\documentclass[11pt,a4paper]{article}

\usepackage{jheppub}
\usepackage{graphicx}
\usepackage{dcolumn}
\usepackage{bm}
\usepackage{amsmath}
\usepackage{braket}
\usepackage{slashed}
\usepackage{epstopdf}
\usepackage{placeins}
\usepackage{multirow}
\usepackage{makecell}
\usepackage{soul}
\usepackage{braket}
\usepackage[shortlabels]{enumitem}
\usepackage{placeins}
\usepackage{subfig}
\usepackage{subcaption}
\usepackage[skins]{tcolorbox}

\usepackage[labelfont=bf, textfont=it, font=small]{caption}
\newcommand{\beq}{\begin{eqnarray}}
\newcommand{\eeq}{\end{eqnarray}}
\newcommand{\beqnn}{\begin{eqnarray*}}
\newcommand{\eeqnn}{\end{eqnarray*}}

\renewcommand{\thesection}{\arabic{section}}

\newcommand{\Tr}{\ensuremath{\mathrm{Tr}}}
\newcommand{\YM}{{\scriptscriptstyle{\mathrm{YM}}}}
\newcommand{\cool}{{\scriptscriptstyle{\mathrm{cool}}}}
\newcommand{\SU}{\ensuremath{\mathrm{SU}}}
\newcommand{\QCD}{{\scriptscriptstyle{\mathrm{QCD}}}}
\newcommand{\W}{{\scriptscriptstyle{\mathrm{W}}}}
\newcommand{\clov}{\scriptscriptstyle{\mathrm{clov}}}
\renewcommand{\L}{{\scriptscriptstyle{\mathrm{L}}}}
\newcommand{\Nr}{N_{\scriptscriptstyle{\mathrm{r}}}}
\newcommand{\Nf}{N_{\scriptscriptstyle{\mathrm{f}}}}
\newcommand{\Ld}{L_{\scriptscriptstyle{\mathrm{d}}}}
\newcommand{\elld}{\ell_{\scriptscriptstyle{\mathrm{d}}}}
\newcommand{\Rs}{R_{\scriptscriptstyle{\mathrm{s}}}}
\newcommand{\dd}{\ensuremath{\mathrm{d}}}

\renewcommand{\inf}{\scriptscriptstyle{\mathrm{\infty}}}

\begin{document}
	
\title{\centering The large-$N$ limit of the topological susceptibility\\of $\mathrm{SU}(N)$ Yang--Mills theories\\via Parallel Tempering on Boundary Conditions}

\author[a]{Claudio Bonanno}

\affiliation[a]{Instituto de F\'isica Te\'orica UAM-CSIC, c/ Nicol\'as Cabrera 13-15, Universidad Aut\'onoma de Madrid, Cantoblanco, E-28049 Madrid, Spain}

\emailAdd{claudio.bonanno@csic.es}

\abstract{I present a large-$N$ determination of the topological susceptibility $\chi$ of SU$(N)$ Yang--Mills theories using non-perturbative numerical Monte Carlo simulations of the lattice-discretized theory for $3\le N \le 6$, and adopting the Parallel Tempering on Boundary Conditions (PTBC) algorithm to bypass topological freezing for $N>3$. Thanks to this algorithm I am able to explore a uniform range of lattice spacings across all values of $N$, and to precisely determine $\chi$ for finer lattice spacings compared to previous studies with periodic or open boundary conditions. By taking the continuum limit at fixed smoothing radius in physical units, I am also able to show the independence of the continuum limit of $\chi$ from this choice. I conclude providing a comprehensive comparison of my new PTBC results with previous determinations of the topological susceptibility in the literature, both at finite $N$ and in the large-$N$ limit.}


\maketitle

\section{Introduction}

Non-abelian $\SU(N)$ Yang--Mills theories in four space-time dimensions possess non-trivial topological properties whose relevance envelops both theoretical sides of gauge theories, and phenomenological aspects of the Standard Model and beyond.

A paramount example is provided by the Witten--Veneziano mechanism~\cite{Witten:1979vv, Veneziano:1979ec}, explaining how the non-trivial topological properties of the QCD vacuum allow the $\eta^\prime$ meson to acquire a non-vanishing mass through the axial anomaly, and how this particle eventually becomes a proper Nambu--Goldstone boson in the limit of infinite number of colors $N\to\infty$. More precisely, taking the large-$N$ limit \emph{\'a la} 't Hooft~\cite{tHooft:1973alw} keeping $\lambda=g^2 N$ fixed, the Witten--Veneziano equation
\beq\label{eq:witten_veneziano_formula}
m^2_{\eta^\prime} = \frac{2 \Nf}{F_\pi^2} \, \chi
\eeq
relates the squared $\eta^\prime$ mass in the chiral limit with $\Nf$ massless quark species to the large-$N$ limit of the topological susceptibility $\chi$ of pure Yang--Mills theories:
\beq
\label{eq:topsusc_continuum}
\chi &=& \lim_{V\to\infty}\frac{\braket{Q^2}}{V},
\eeq
with $Q$ the integer-valued topological charge
\beq
\label{eq:topcharge_continuum}
Q &=& \frac{1}{32\pi^2}\varepsilon_{\mu\nu\rho\sigma} \int \dd^4 x \, \Tr\left[G_{\mu\nu}(x)G_{\rho\sigma}(x)\right] \in \mathbb{Z}.
\eeq

Since $m^2_{\eta^\prime}\sim 1/N$ and the pion decay constant $F^2_\pi\sim N$, Eq.~\eqref{eq:witten_veneziano_formula} requires $\chi$ to approach a finite large-$N$ limit. Due to the crucial importance of the large-$N$ behavior of $\chi$ for the theoretical understanding of both the chiral anomaly and the $\eta^\prime$ phenomenology, this quantity has been the object of several studies. Since topological properties are purely non-perturbative aspects of gauge theories, and analytical tools such as effective chiral Lagrangians~\cite{DiVecchia:1980yfw, diCortona:2015ldu} or semiclassical arguments~\cite{Gross:1980br,Schafer:1996wv,Boccaletti:2020mxu} are capable of providing reliable results only in certain regimes and under certain assumptions, numerical Monte Carlo (MC) simulations of lattice-discretized pure-gauge theories provide the most natural framework to address the determination of $\chi$ in the large-$N$ limit from first principles.

On general grounds, lattice simulations have proven to be a reliable tool to study large-$N$ gauge theories, and have been extensively employed to this end~\cite{DelDebbio:2001sj,Lucini:2001ej,Lucini:2004my,DelDebbio:2006yuf,Vicari:2008jw,Allton:2008ty,Lucini:2010nv,Lucini:2012gg,Bali:2013kia,Bonati:2016tvi,Ce:2016awn,Bennett:2020hqd,Athenodorou:2021qvs,Bennett:2022gdz,Athenodorou:2024loq,Sharifian:2025fyl, DeGrand:2016pur,Hernandez:2019qed,DeGrand:2020utq,Hernandez:2020tbc,DeGrand:2021zjw,Baeza-Ballesteros:2022azb,DeGrand:2023hzz,DeGrand:2024lvp,DeGrand:2024frm,Baeza-Ballesteros:2025iee,Butti:2025rnu, PhysRevLett.48.1063,BHANOT198247,Gross:1982at,GONZALEZARROYO1983174,PhysRevD.27.2397,Aldazabal:1983ec,Kiskis:2002gr,Narayanan:2003fc,Kovtun:2007py,Unsal:2008ch,Gonzalez-Arroyo:2010omx,Neuberger:2020wpx, Gonzalez-Arroyo:1983cyv,Das:1984jh,Das:1985xc,Narayanan:2004cp,Gonzalez-Arroyo:2005dgf,Kiskis:2009rf,Hietanen:2009ex,Hietanen:2010fx,Bringoltz:2011by,Hietanen:2012ma,Gonzalez-Arroyo:2012euf,Gonzalez-Arroyo:2013bta,Lohmayer:2013spa,Gonzalez-Arroyo:2014dua,GarciaPerez:2014azn,GarciaPerez:2015rda,Perez:2015ssa,Gonzalez-Arroyo:2015bya,Perez:2017jyq,GarciaPerez:2020gnf,Perez:2020vbn,Butti:2023hfp,Bonanno:2023ypf,Butti:2022sgy,Bonanno:2024bqg,Bonanno:2024onr,Hamada:2025whg,Bonanno:2025hzr}. However, the lattice investigation of gauge topology in the large-$N$ limit is known to be a challenging task due to a serious algorithmic limitation: the infamous \emph{topological freezing} problem. When adopting standard local updating algorithms to sample the lattice Yang--Mills path integral, the autocorrelation time of $Q$, i.e., the number of MC steps necessary to draw two decorrelated occurrences of the topological charge, rapidly diverges as function of the inverse lattice spacing $a$ as the continuum limit $a\to 0$ is approached~\cite{Alles:1996vn,DelDebbio:2002xa,DelDebbio:2004xh,Schaefer:2010hu}. This is understood in terms of the Monte Carlo dynamics of topological modes: achieving a jump of the topological charge of the lattice gauge configuration via a local field deformation becomes increasingly more difficult as the lattice spacing gets finer due to the development of potential barriers among pseudo-topological sectors, which eventually diverge in the continuum limit to restore a proper notion of integer-valued winding number~\cite{Luscher:1981zq}. The rapid growth of the autocorrelation time of the topological charge thus induces a severe loss of ergodicity of the Monte Carlo Markov chain, which tends to remain frozen in a fixed topological sector for a very long time, preventing a correct sampling of the topological charge distribution. Moreover, increasing the rank of the gauge group $N$ at fixed lattice spacing further worsens the severity of freezing: in practice, at large $N$ topological freezing kicks in already at coarse lattice spacing, making it very challenging to study the continuum limit of topological quantities in this regime.

Since the loss of ergodicity due to topological freezing can in principle bias any expectation value computed from MC simulations~\cite{Brower:2003yx, Aoki:2007ka}, the impact of this computational problem goes beyond the mere investigation of gauge topology itself. As a matter of fact, the last decade saw an intense strive to develop novel algorithmic strategies to mitigate its severity, see, e.g.,~\cite{Bietenholz:2015rsa,Laio:2015era,Luscher:2017cjh,Bonati:2017woi,Giusti:2018cmp,Florio:2019nte,Funcke:2019zna,Kanwar:2020xzo,Nicoli:2020njz,Albandea:2021lvl,Cossu:2021bgn,Borsanyi:2021gqg,papamakarios2021,Fritzsch:2021klm,Abbott:2023thq,Eichhorn:2023uge,Howarth:2023bwk,Albandea:2024fui,Bonanno:2024udh,Vadacchino:2024lob,Abe:2024fpt} (see also the recent reviews~\cite{Finkenrath:2023sjg,Boyle:2024nlh,Finkenrath:2024ptc}). Among these, a novel algorithmic proposal that has proven to be particularly effective in mitigating topological freezing, allowing to advance the state of the art about the investigation of gauge topology on the lattice in several directions, is the \emph{Parallel Tempering on Boundary Conditions} (PTBC). This algorithm, initially proposed~\cite{Hasenbusch:2017unr} and applied~\cite{Berni:2019bch} to $2d$ CP$^{N-1}$ models (see also~\cite{Bonanno:2022hmz}), has by now been extensively applied also to $4d$ non-abelian gauge theories, both in the pure-gauge case~\cite{Bonanno:2020hht,Bonanno:2022yjr,Bonanno:2023hhp,Bonanno:2024nba,Bonanno:2024ggk} and in full QCD with dynamical fermions and physical quark masses~\cite{Bonanno:2024zyn}, revealing itself as a versatile tool to efficiently explore fine lattice spacings and large values of $N$ both at vanishing and non-zero temperatures.\\

The main idea behind this algorithm is to perform simultaneous simulations of several lattice replicas, each one enjoying different boundary conditions, within a parallel tempering framework. This means that each replica is updated independently using standard methods, and is allowed to swap its gauge field configuration with neighboring ones via a standard accept/reject Metropolis step. The boundary conditions of the replicas are chosen so as to interpolate between Periodic Boundary Conditions (PBC) and Open Boundary Conditions (OBC). As it is well known, topological freezing is drastically mitigated in the presence of OBC~\cite{Luscher:2011kk}, since potential barriers among pseudo-topological sectors are removed, and topological fluctuations are injected/ejected from the boundaries and then diffuse around the lattice~\cite{McGlynn:2014bxa}. However, this does not come for free: OBC simulations are affected by enhanced finite-volume effects, as physical fluctuations are only found in the bulk of the lattice, sufficiently far from the boundaries. Moreover, in the presence of OBC translation invariance and thus a proper definition of global topological charge are lost. The tempering on the boundary conditions is exactly introduced to take advantage of the improved autocorrelation of $Q$ achieved with open boundaries, while at the same time avoiding systematic boundary effects. The swaps among replicas diffuse topological fluctuations created on the OBC replica towards the PBC one, which in this way enjoys the improved autocorrelations of $Q$. At the same time, the PBC replica keeps translation invariance, thus, it is selected as the system where all physical observables are computed free of systematic boundary effects. Clearly, the necessity of simulating more systems at once increases the computational burden by a factor equal to the number of replicas, but this is largely compensated by the drastic reduction of autocorrelations and finite-size boundary effects, leading in the end the PTBC algorithm to outperform standard ones by up to several orders of magnitude~\cite{Hasenbusch:2017unr,Berni:2019bch,Bonanno:2022hmz,Bonanno:2020hht,Bonanno:2022yjr,Bonanno:2023hhp,Bonanno:2024nba,Bonanno:2024ggk}.

While this algorithm has allowed the determination of several quantities whose calculation was either unsatisfactorily or at all attempted previously with other algorithms~\cite{Hasenbusch:2017unr,Berni:2019bch,Bonanno:2022hmz,Bonanno:2020hht,Bonanno:2022yjr,Bonanno:2023hhp,Bonanno:2024nba,Bonanno:2024ggk}, so far only a few determinations of $\chi$ using the PTBC algorithm were provided. Some for $N=4,6$ can be found in Ref.~\cite{Bonanno:2020hht}, but no dedicated study of the susceptibility was performed in that paper. The goal of the present investigation is to fill this gap. I will present a dedicated high-statistics lattice study of the topological susceptibility at large $N$ using the PTBC algorithm to effectively mitigate topological freezing. Thanks to PTBC, I will simulate Yang--Mills theories for $N=4,5,6$ across the same range of lattice spacings avoiding topological freezing. The probed parameter space will include finer lattice spacings compared to previous investigations with PBC~\cite{Bonati:2015sqt,Bonati:2016tvi,Athenodorou:2020ani,Athenodorou:2021qvs} and OBC~\cite{Ce:2015qha,Ce:2016awn}, and will be similar in extension to the one used in typical $N=3$ studies. I will then combine these new large-$N$ results with the high-statistics ones I obtained for $N=3$ in~\cite{Bonanno:2023ple} to study the large-$N$ limit of $\chi$. Finally, I will conclude with a critical comparison of the determinations of $\chi(N)$ presented in this study with earlier ones obtained with different methods and algorithms.

This paper is organized as follows: I will describe the employed numerical setup and the implementation of the PTBC algorithm in Sec.~\ref{sec:setup}; I will present my numerical results for the topological susceptibility in Sec.~\ref{sec:res}; finally, I will draw my conclusions in Sec.~\ref{sec:conclu}.

\section{Numerical setup}\label{sec:setup}

In this section I will describe the employed numerical setup, namely, the implementation of the PTBC algorithm, the adopted lattice action, and the numerical strategy followed to determine the topological susceptibility $\chi$.

\subsection{Lattice action and PTBC algorithm}

The pure Yang--Mills action
\beq\label{eq:ym_action_cont}
\mathcal{S}_{\YM} &=& \frac{N}{2\lambda}\int \dd^4 x \, \Tr\left[G_{\mu\nu}(x)G_{\mu\nu}(x)\right], \qquad \lambda=g^2 N,
\eeq
is discretized on a hyper-cubic space-time lattice with $L$ points for each side and lattice spacing $a$ using the standard Wilson plaquette action. In the presence of periodic boundary conditions in all directions, the lattice action reads:
\beq\label{eq:lat_action_periodic}
\mathcal{S}_{\W}[U] &=& -N b \sum_{x} \sum_{\mu>\nu} \Re\,\Tr \, U_{\mu\nu}(x),\\
U_{\mu\nu}(x) &=& U_\mu(x)U_{\nu}(x+a\hat{\mu})U^\dagger_\mu(x+a\hat{\nu})U^\dagger_\nu(x),
\eeq
where $U_\mu(x) \in \SU(N)$ are the gauge link variables, $U_{\mu\nu}(x)$ is the plaquette in the site $x$ lying on the $(\mu,\nu)$ plane, and $b=1/\lambda=\beta/(2N^2)$ is the inverse bare 't Hooft coupling. The other replicas are introduced by altering the boundary conditions on a small sub-region of the lattice, the \emph{defect}, which is here chosen to be a cubic region of size $(a\Ld)^3$ placed on the temporal boundary. This is in practice achieved by altering the lattice action in Eq.~\eqref{eq:lat_action_periodic} as follows:
\beq
\mathcal{S}_{\W}[U,c] = -N b \sum_{x} \sum_{\mu>\nu} \mathcal{K}^{(c)}_{\mu\nu}(x)\, \Re\,\Tr \, U_{\mu\nu}(x)
\eeq
where the factor
\beq
\mathcal{K}^{(c)}_{\mu\nu}(x) = K^{(c)}_{\mu}(x) K^{(c)}_{\nu}(x+a\hat{\mu}) K^{(c)}_{\mu}(x+a\hat{\nu}) K^{(c)}_{\nu}(x),
\eeq
\beq
K^{(c)}_{\mu}(x) =
\begin{cases}
c,  & \mu=0\,,\,\, x \in D\, ,\\
1,     & \text{otherwise,}
\end{cases}
\eeq
suppresses the coupling of the links in the temporal direction crossing the defect region $D=\{x_0=a(L-1), 0\le x_1,x_2,x_2 \le a\Ld\}$ by a replica-dependent factor of $c$. The replica dependent factor $0\le c(r) \le 1$, with $r$ the replica index, is used to interpolate between PBC, $c(r=0)=1$, and OBC, $c(r=\Nr-1)=0$, with $\Nr$ the number of replicas. Indeed, when $c=1$, one gets back the periodic action~\eqref{eq:lat_action_periodic}. When $c=0$, the gauge coupling of the affected links vanishes; they are thus kept fixed and not updated during the MC evolution, realizing OBC. 

Each replica is updated using standard methods: an elementary MC updating step consists of 1 lattice sweep of heat-bath~\cite{Creutz:1980zw,Kennedy:1985nu} followed by 4 lattice sweeps of over-relaxation~\cite{Creutz:1987xi}, both implemented \emph{\'a la} Cabibbo--Marinari~\cite{Cabibbo:1982zn}, i.e., updating all the $N(N-1)/2$ $\SU(2)$ subgroups of $\SU(N)$. At the end of each updating step, swaps among lattice gauge configurations are proposed among all adjacent replicas $r, r^\prime=r+1$, and are accepted/rejected according to the Metropolis probability:
\beq
p(r) = \min\big[1, \exp\{-\Delta \mathcal{S}(r)\}\big],
\eeq
with $\Delta \mathcal{S}(r)$ the variation of the total action after the swap of configurations,
\beq
\begin{aligned}
\Delta \mathcal{S}(r) = & \quad \,\mathcal{S}_{\W}[U(r),c(r+1)] + \mathcal{S}_{\W}[U(r+1),c(r)]\\
&- \mathcal{S}_{\W}[U(r),c(r)] - \mathcal{S}_{\W}[U(r+1),c(r+1)].
\end{aligned}
\eeq
The calculation of $\Delta \mathcal{S}(r)$ has a negligible cost being non-vanishing only around $D$. After swaps, updating sweeps of the full lattice are alternated with some hierarchical updates concentrated around the defect $D$, which is the region where most of the new topological excitations are created/annihilated. These are chosen to have a negligible total numerical cost with respect to the full sweeps (more technical details are found in~\cite{Hasenbusch:2017unr,Bonanno:2020hht}). Each hierarchical update is followed by swap proposals too. Finally, after each series of swap proposals, a random translation of one lattice spacing in a random direction is performed on the periodic lattice, which is translation invariant. Indeed, since the region $D$ is kept fixed, performing random translations of the PBC replica effectively moves the position of the defect, so that topological excitations are created/annihilated around the lattice.

The boundary condition coefficients $c(r)$ are tuned through short test runs in order to ensure a uniform swap acceptance rate $p(r) \approx p$, which is the optimal choice for parallel tempering simulations, as it ensures that the field configurations move uniformly among the replicas in a random walk fashion. All simulations will be performed fixing the uniform acceptance rate $p$ and the size of the defect in physical units $\elld = a\Ld$. Once these two quantities are kept fixed, the number of replicas $\Nr$ becomes just a function of the lattice spacing $a$ and of the number of colors $N$. More details on the adopted choices for $p$, $\Nr$ and $\elld$ will be given later in Sec.~\ref{sec:simul_params}.

The PTBC code implementation I have used for this investigation can be found in~\cite{PTBC}.

\subsection{Determination of the topological susceptibility}\label{sec:setup_topsusc}

Given that I will be using the PTBC algorithm throughout this study, I will always compute observables in the periodic replica, where translation invariance is kept, and where a global topological charge can be defined, thus allowing the use of Eq.~\eqref{eq:topsusc_continuum} to compute the topological susceptibility. I will discretize the topological charge in Eq.~\eqref{eq:topcharge_continuum} adopting a standard gluonic definition, the clover topological charge:
\beq
Q_{\clov} = \frac{1}{32\pi^2} \sum_x \sum_{\mu\nu\rho\sigma}\varepsilon_{\mu\nu\rho\sigma}\Tr\left[C_{\mu\nu}(x)C_{\rho\sigma}(x)\right],
\eeq
\beq
C_{\mu\nu}(x) = \frac{1}{4} \Im\left\{U_{\mu\nu}(x) + U_{-\nu, \mu}(x) + U_{\nu, -\mu}(x) + U_{-\mu, -\nu}(x)\right\},
\eeq
where $C_{\mu\nu}(x)$ is the four-leaf clover, and where $U_{-\mu}(x) = U^\dagger_{\mu}(x-a\hat{\mu})$ is understood. This is the simplest parity-odd discretization of Eq.~\eqref{eq:topcharge_continuum}.

It is well-known that gluonic discretizations are affected by multiplicative renormalizations~\cite{Campostrini:1988cy,Vicari:2008jw} that make them non-integer on the lattice:
\beq
Q_{\clov} = Z Q,
\eeq
with $Z \to 1$ in the continuum limit $a \to 0$. Moreover, when used to compute integrated correlation functions, such as the topological susceptibility, 
\beq
\chi = \frac{1}{V} \braket{Q^2} = \int \dd^4 x \, \braket{q(x)q(0)},
\eeq
\beq
Q=\int \dd^4 x \, q(x), \qquad q(x) = \frac{1}{32\pi^2} \varepsilon_{\mu\nu\rho\sigma}\Tr\left[G_{\mu\nu}(x)G_{\rho\sigma}(x)\right],
\eeq
they develop contact terms resulting in further additive renormalizations~\cite{DiVecchia:1981aev,DElia:2003zne,Vicari:2008jw}:
\beq
\chi_{\clov}=\frac{1}{V}\braket{Q^2_{\clov}} = Z^2 \chi + M.
\eeq
Such additive terms eventually diverge in the continuum limit, $M\to \infty$ when $a \to 0$, thus overcoming the physical signal.

A customary strategy to deal with these renormalizations is to employ smoothing algorithms. Smoothing is used to dump ultraviolet (UV) fluctuations and drive a lattice gauge configuration towards the nearest local minimum. If not overly prolonged, it is expected to leave the global topological background of a gauge configuration unaltered, being this an infrared (IR) feature of the fields. Another important aspect of smoothing algorithms is that they act as diffusive processes, and kill short-distance fluctuations below a UV scale called \emph{smoothing radius} $\Rs$, proportional to the square root of the amount of smoothing performed. Thus, after smoothing, the lattice topological charge will exhibit a distribution peaked around integer values, and when used to compute the topological susceptibility, will yield a properly renormalized definition with $Z\simeq 1$ and $M\simeq0$, which will scale towards the continuum limit to the correct physical quantity without short-distance singularities~\cite{Ce:2015qha,Ce:2016awn}.

Several smoothing algorithms have been proposed and employed in the literature, such as cooling~\cite{Berg:1981nw,Iwasaki:1983bv,Itoh:1984pr,Teper:1985rb,Ilgenfritz:1985dz,Campostrini:1989dh,Alles:2000sc}, gradient flow~\cite{Narayanan:2006rf,Luscher:2009eq, Luscher:2010iy,Lohmayer:2011si} or stout smearing~\cite{APE:1987ehd, Morningstar:2003gk}. These smoothing methods have been shown to be fully consistent among each other when smoothing radii are matched across different procedures, see~\cite{Alles:2000sc, Bonati:2014tqa, Alexandrou:2015yba}. In this paper, I will use cooling for its simplicity and numerical cheapness. One cooling step consists in a lattice sweep where each gauge link is iteratively aligned to its relative local staple in order to locally minimize the Wilson action~\eqref{eq:lat_action_periodic}. By matching with the Wilson flow, it has been shown that for cooling the smoothing radius in lattice units is given by~\cite{Bonati:2014tqa}:
\beq\label{eq:smooth_radius}
\frac{\Rs}{a} = \sqrt{\frac{8}{3} n_\cool} = \sqrt{8 \frac{t}{a^2}},
\eeq
with $n_{\cool}$ the number of cooling steps and $t/a^2=n_{\cool}/3$ its equivalent gradient flow time.

\noindent Denoting with $Q_{\L}(\Rs)$ the clover topological charge computed after $n_{\cool}=\frac{3}{8}\left(\frac{\Rs}{a}\right)^2$ cooling steps, I will thus compute the lattice topological susceptibility as:
\beq
\chi_{\L}(\Rs) = \frac{1}{V} \braket{Q^2_{\L}(\Rs)}, \qquad V=(aL)^4.
\eeq
In the continuum limit, the topological susceptibility is expected to be independent of the smoothing radius $\Rs$~\cite{Ce:2015qha,Ce:2016awn,Giusti:2018cmp,Bonanno:2023ple,Durr:2025qtq}, at least if this is taken sufficiently small so as to achieve an effective separation between this scale and the one characterizing the relevant long-distance fluctuations dominantly contributing to $\chi$. However, the choice of the smoothing radius will of course influence the value of the lattice susceptibility, thus, the expected leading continuum scaling of $\chi_{\L}(\Rs)$ is:
\beq
\chi_{\L}(a,\Rs) = \chi + k(\Rs) a^2 + o(a^2),
\eeq
i.e., one expects that the  value of $\Rs$ should only influence the slope of $\chi_{\L}$ as a function of $a^2$, but not its continuum limit $\chi$. In the following, I will verify this behavior by taking the continuum limit of $\chi_{\L}$ at fixed value of $\Rs$ in physical units for several values of $\Rs$.

\section{Numerical results}\label{sec:res}

This section is devoted to the presentation of the new PTBC results for the topological susceptibility. I will start by discussing the explored parameter space and the details of the PTBC algorithm, the obtained autocorrelation times and the adopted scale setting. Then, I will present the continuum and large-$N$ extrapolations of the obtained lattice results for $\chi$. Finally, I will compare my new determinations with previous ones in the literature.

\subsection{Simulation parameters, autocorrelation times and scale setting}\label{sec:simul_params}

\begin{table}[!t]
\small
\begin{center}
\begin{tabular}{|c|c|c|c|c|c|c|c|c|c|c|c|}
\hline
&&&&&&&&&&&\\[-1em]
$N$ & $\beta$ & $b=\dfrac{1}{\lambda}$ & $L$ & $a\sqrt{\sigma}$ & $\ell\sqrt{\sigma}$ & \makecell{$n_\cool$\\max} & \makecell{$\Rs \sqrt{\sigma}$\\max} & $n_{\scriptscriptstyle{\rm meas}}$ & $\Nr$ & $\Ld$ & $\elld \sqrt{\sigma}$\\
\hline
\multicolumn{12}{c}{}\\[-1em]
\cline{1-9}
\multirow{5}{*}{3} & 5.95 & 0.3306 & 16 & 0.23567(69) & 3.77 & 19 & 1.678 & 1.94M & \multicolumn{3}{c}{} \\
& 6.00 & 0.3333 & 16 & 0.21609(76) & 3.46 & 22 & 1.654 & 2.03M & \multicolumn{3}{c}{} \\
& 6.07 & 0.3372 & 18 & 0.19238(51) & 3.46 & 28 & 1.662 & 1.94M & \multicolumn{3}{c}{} \\
& 6.20 & 0.3444 & 22 & 0.15788(31) & 3.47 & 42 & 1.668 & 1.23M & \multicolumn{3}{c}{} \\
& 6.40 & 0.3556 & 30 & 0.11879(26) & 3.57 & 75 & 1.682 & 0.71M & \multicolumn{3}{c}{} \\
\cline{1-9}
\multicolumn{12}{c}{}\\[-1em]
\cline{1-9}
\multirow{4}{*}{4} & 11.02 & 0.3444 & 16 & 0.21434(28) & 3.43 & 21 & 1.679 & 2.08M & \multicolumn{3}{c}{} \\
\cline{10-12}
& 11.20 & 0.3500 & 20 & 0.18149(49) & 3.63 & 32 & 1.677 & 198k & 16 & 3 & 0.54 \\
& 11.40 & 0.3563 & 24 & 0.15305(34) & 3.67 & 45 & 1.677 & 206k & 16 & 3 & 0.46 \\
& 11.60 & 0.3625 & 26 & 0.13065(21) & 3.40 & 64 & 1.706 & 262k & 24 & 4 & 0.52 \\
\hline
\multicolumn{12}{c}{}\\[-1em]
\hline
\multirow{4}{*}{5} & 17.43  & 0.3486 & 16 & 0.22217(37) & 3.55 & 21 & 1.663 & 777k & 12 & 2 & 0.44 \\
& 17.63  & 0.3526 & 18 & 0.19636(35) & 3.53 & 28 & 1.697 & 464k & 20 & 3 & 0.59 \\
& 18.04  & 0.3608 & 22 & 0.15622(38) & 3.44 & 42 & 1.653 & 380k & 20 & 3 & 0.47 \\
& 18.375 & 0.3675 & 26 & 0.13106(30) & 3.41 & 60 & 1.658 & 118k & 32 & 4 & 0.52 \\
\hline
\multicolumn{12}{c}{}\\[-1em]
\hline
\multirow{4}{*}{6} & 25.32 & 0.3517 & 16 & 0.22208(35) & 3.55 & 20 & 1.622 & 611k & 14 & 2 & 0.44 \\
& 25.70 & 0.3569 & 18 & 0.18956(33) & 3.41 & 28 & 1.638 & 325k & 24 & 3 & 0.57 \\
& 26.22 & 0.3642 & 22 & 0.15480(36) & 3.41 & 42 & 1.638 & 200k & 24 & 3 & 0.46 \\
& 26.65 & 0.3701 & 26 & 0.13173(29) & 3.42 & 57 & 1.624 & 115k & 36 & 4 & 0.53 \\
\hline
\end{tabular}
\end{center}
\caption{Summary of simulation parameters for $N=4,5,6$, along with those of the $N=3$ calculations of~\cite{Bonanno:2023ple}. All simulations for $N>3$, except for the coarsest lattice spacing at $N=4$, were done adopting the PTBC algorithm with an average swap acceptance $p\approx 20\%$. Scale setting is performed in terms of the string tension $\sigma$, according to a cubic spline interpolation of the determinations of Refs.~\cite{Athenodorou:2020ani,Athenodorou:2021qvs}. For each point I report both the standard inverse coupling $\beta=2N/g^2$ and the inverse 't Hooft coupling $b=1/\lambda=1/(Ng^2)$ (with 4 significant digits). The quantity $n_{\scriptscriptstyle{\rm meas}}$ is the number of samples of $Q_\L$, always measured every 10 updating steps (except for $\beta=6.40$, where measures are separated by 100 updates).}
\label{tab:params}
\end{table}

All simulation parameters of the new calculations done in this study for $N=4,5,6$ are reported in Tab.~\ref{tab:params}, along with those of the $N=3$ calculations I did in Ref.~\cite{Bonanno:2023ple}. I have explored a uniform range of lattice spacings, corresponding to $0.22 \gtrsim a\sqrt{\sigma} \gtrsim 0.13$, with $\sigma$ the string tension. This is the chosen scale setting quantity since it is the only one I could retrieve covering the full parameter space I explored. In previous studies~\cite{Bonati:2015sqt,Bonati:2016tvi,Ce:2015qha,Ce:2016awn,Bonanno:2023ple} it has been shown that choosing the physical size of the lattice $\ell = aL$ above $\sim 3/\sqrt{\sigma}$ (about 1.33 fm assuming $\sqrt{\sigma}=445$ MeV~\cite{Bulava:2024jpj}) is sufficient to avoid finite-size effects, thus I have chosen an approximately fixed lattice size $\ell\sqrt{\sigma} \sim 3.4-3.7$ in all cases.

All $N>3$ simulations, except for the coarsest lattice spacing of $N=4$ (being autocorrelations still small in that case), were performed adopting the PTBC algorithm, choosing a uniform acceptance rate of $p\approx 20\%$, and an approximately fixed defect size $\elld\sqrt{\sigma} \sim 0.45-0.55$ (about $0.2-0.25$ fm assuming $\sqrt{\sigma}=445$ MeV~\cite{Bulava:2024jpj}), i.e., about $\sim 13-15\%$ of the total lattice size. This choice for $\elld$ was shown~\cite{Bonanno:2020hht,Bonanno:2024nba,Bonanno:2024ggk,Bonanno:2024zyn} to be optimal with the purpose of minimizing the following figure of merit:
\beq\label{eq:figure_merit_tau}
\tau(Q^2) = \Nr \tau_0(Q^2),
\eeq
within the range $\elld\sqrt{\sigma} \sim 0.2 - 0.85$. In Eq.~\eqref{eq:figure_merit_tau}, $\tau_0(Q^2)$ stands for the PTBC (integrated) autocorrelation time of the squared lattice topological charge expressed in units of the MC time of the periodic system, while $\tau(Q^2)$ represents the same quantity but in units of the number of standard MC updating steps (i.e., taking into account the replica overhead). This is a useful quantity to monitor performances, as a larger defect would decrease $\tau_0$, but would also increase the required number of replicas $\Nr$ to keep the same swap acceptance. Varying $\elld\sqrt{\sigma}$ between $0.2$ and $0.85$ it was found in~\cite{Bonanno:2020hht,Bonanno:2024nba,Bonanno:2024ggk,Bonanno:2024zyn} that $\elld\sqrt{\sigma}\sim 0.45-0.55$ yielded the best compromise between these two effects, achieving the minimum $\tau$. Similarly, it was also shown in~\cite{Bonanno:2020hht,Bonanno:2024nba,Bonanno:2024ggk,Bonanno:2024zyn} that varying the swap acceptance parameter $p$ in the range $\sim 20-30\%$ gave perfectly compatible results for $\tau(Q^2)$ for $\elld\sqrt{\sigma}\sim 0.45-0.55$, as increasing $p$ was found in~\cite{Bonanno:2020hht,Bonanno:2024nba,Bonanno:2024ggk,Bonanno:2024zyn} to decrease $\tau_0$ and increase $\Nr$ by the same amount, yielding compatible values of $\tau$. Thus, I opted for $p\approx 20\%$ to use a smaller number of replicas. Within the explored range of $a$, the desired $\elld$ is obtained choosing $\Ld$ of the order of $2-4$ lattice spacings, and I could reach $p\approx 20 \%$ with $\Nr \sim \mathcal{O}(10)$.

\newpage

\begin{figure}[!t]
\centering
\includegraphics[scale=0.45]{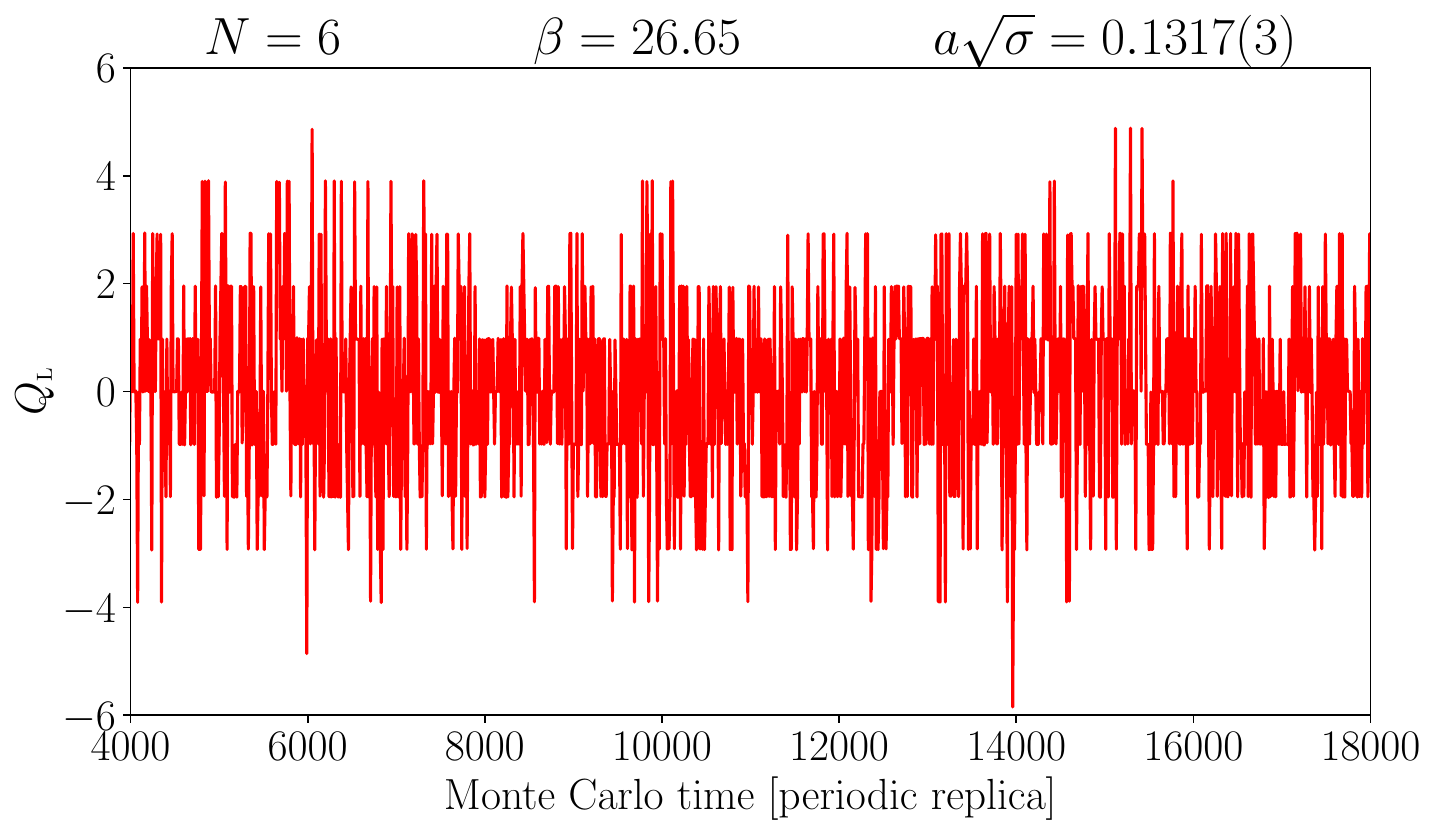}
\caption{Example of the Monte Carlo evolution of the cooled topological charge $Q_\L$ for $\Rs\sqrt{\sigma}\simeq 1.18$ for the finest lattice spacing and largest $N$ explored in this study.}
\label{fig:history_Q_example}
\end{figure}

The efficiency of PTBC simulations is clearly spelled out by Fig.~\ref{fig:history_Q_example}, where I reported an example of the Monte Carlo evolution of the lattice topological charge for the finest lattice spacing simulated at the largest value of $N$ for a small time window (corresponding to $\sim 10 \%$ of the whole generated sample). The plot just refers to the Monte Carlo time of the periodic replica, meaning that it is expressed in units of standard updating steps (equal to 1 lattice sweep of heat-bath and 4 lattice sweeps of over-relaxation). As it can be observed, the PTBC algorithm is very effective in obtaining a rapidly fluctuating total topological charge on the lattice with periodic boundaries, resulting in a Monte Carlo sample with small autocorrelations. Indeed, with the adopted choices of the PTBC parameters, the autocorrelation time of the squared topological charge $\tau_0(Q^2)$ in the periodic system always turned out to be of the order of a few tens, see Tab.~\ref{tab:autocorrs}, with a mild dependence on $N$ and on $a$ in the explored parameter space (see below for further details on this point). Thus, measuring observables every 10 updating steps was in all cases sufficient to reliably estimate statistical errors via a standard binned jack-knife analysis.

\begin{table}[!t]
\begin{center}
\begin{tabular}{|c|c|c|c|c|c|}
\hline
&&&&&\\[-1em]
$N$ & $\beta$ & $a\sqrt{\sigma}$ & $\Nr$ & $\tau_0(Q^2)$ & $\tau(Q^2) = \Nr \tau_0(Q^2)$\\
\hline
\hline
\multirow{3}{*}{4} & 11.20  & 0.18149(49) & 16 & 12(2) & 192(32) \\
& 11.40  & 0.15305(34) & 16 & 35(6)  & 560(96) \\
& 11.60  & 0.13065(21) & 24 & 41(11) & 984(264) \\
\hline
\hline
\multirow{4}{*}{5} & 17.43  & 0.22217(37) & 12 & 18(2) & 216(24) \\
& 17.63  & 0.19636(35) & 20 & 12(2)  & 240(40) \\
& 18.04  & 0.15622(38) & 20 & 53(14) & 1060(280) \\
& 18.375 & 0.13106(30) & 32 & 36(11) & 1152(352) \\
\hline
\hline
\multirow{4}{*}{6} & 25.32  & 0.22208(35) & 14 & 28(5) & 392(70) \\
& 25.70  & 0.18956(33) & 24 & 13(3)  & 312(72) \\
& 26.22  & 0.15480(36) & 24 & 58(12) & 1392(288) \\
& 26.65  & 0.13173(29) & 36 & 40(12) & 1440(432) \\
\hline
\end{tabular}
\end{center}
\caption{Compilation of $\tau_0(Q^2)$, the integrated autocorrelation time of $Q^2_{\L}(\Rs)$ of the periodic system. It is expressed in units of the Monte Carlo time of the periodic replica and obtained from a standard binned jack-knife analysis. Data refer to the lattice clover charge measured after cooling with $\Rs \sqrt{\sigma}\approx 0.8$. The figure of merit $\tau(Q^2) = \Nr \tau_0(Q^2)$ takes into account the replica overhead.}
\label{tab:autocorrs}
\end{table}

\begin{figure}[!t]
\centering
\includegraphics[scale=0.5]{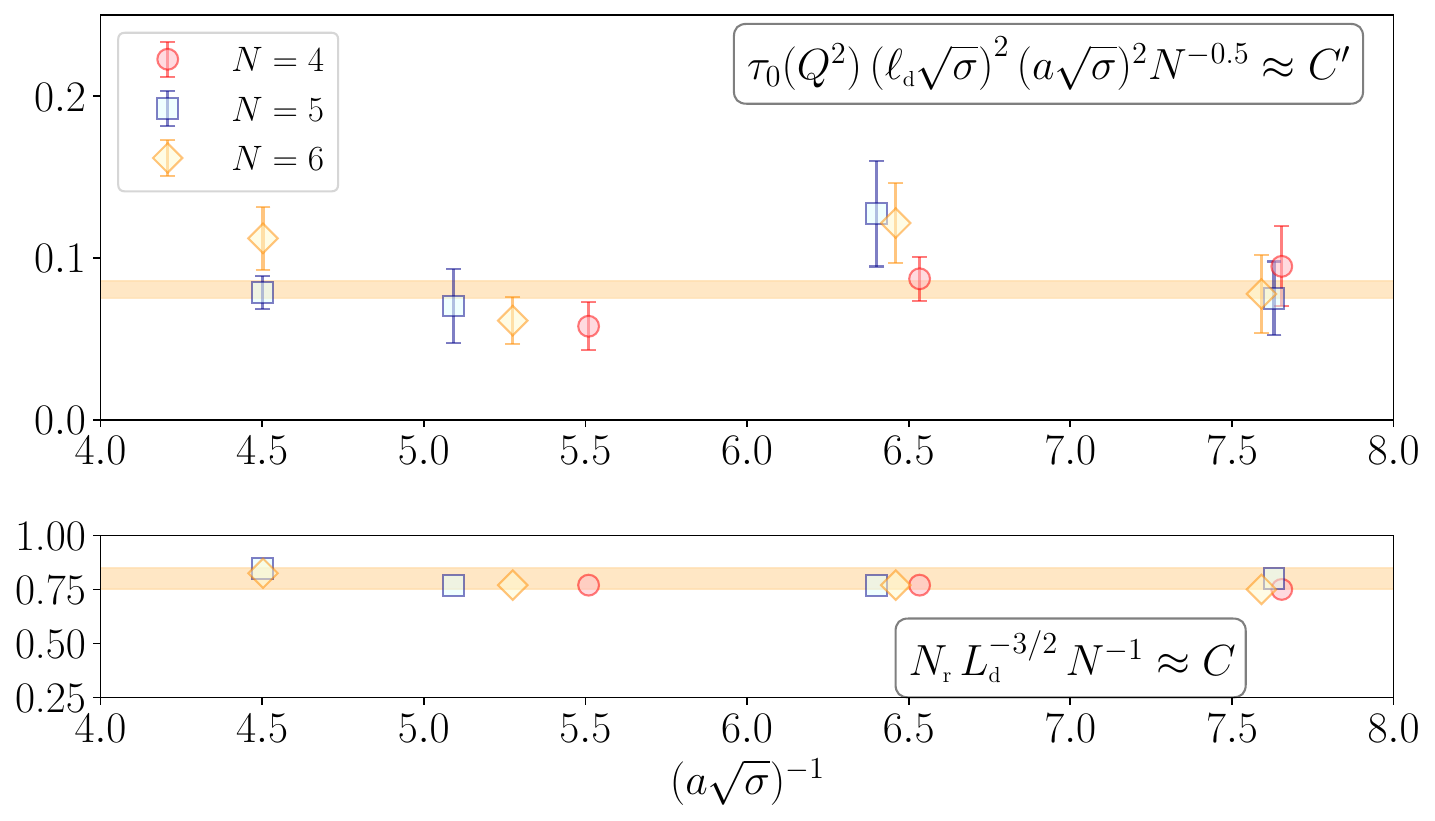}
\caption{Top panel shows the autocorrelation times $\tau_0$ of $Q^2$ obtained with PTBC algorithm on the periodic system as a function of the lattice spacing $a$ and of the rank of the gauge group $N$. The corresponding number of replicas required is shown in the bottom panel. The empirical rescalings reported in these plots have been determined by looking for a collapse of the data around an approximate constant value, see the text for more details.}
\label{fig:autocorrs}
\end{figure}

The probed parameter space also allowed me to assess the scaling of the performances of the PTBC algorithm as a function of $a$ and $N$. As earlier anticipated, at fixed $p$ and defect size $\elld$, the number of replicas $\Nr$ is just a function of the lattice spacing $a$ and of the number of colors $N$. Empirically, I find that:
\beq\label{eq:Nr_scaling}
\Nr \approx C N \Ld^{3/2} = C N \left(\frac{\elld\sqrt{\sigma}}{a\sqrt{\sigma}}\right)^{3/2}, \qquad C \approx 0.80(5),
\eeq
cf.~Fig.~\ref{fig:autocorrs}, confirming the findings of previous PTBC studies for coarser lattice spacings~\cite{Bonanno:2020hht,Bonanno:2024nba,Bonanno:2024ggk}. This is also consistent with a general result about parallel tempering simulations, stating that $\Nr$ should scale as the square root of the number of degrees of freedom interested by the tempering~\cite{Hukushima:1995mcr}, which in this case is $\mathcal{O}(N^2\Ld^3)$.\\
Concerning instead the scaling of autocorrelations, I empirically observe that $\tau_0(Q^2)$ scatters around an approximately constant value if I perform the following rescaling, see Fig.~\ref{fig:autocorrs}:
\beq\label{eq:rescaled_tau0}
\tau_0(Q^2) \longrightarrow \tau_0(Q^2) \left(\elld\sqrt{\sigma}\right)^2 \left(a\sqrt{\sigma}\right)^2 N^{-0.5} \approx C^\prime.
\eeq
Indeed, rescaled $\tau_0(Q^2)$ can be fitted to a constant with a reduced chi-squared of $\sim 1.25$ with 10 degrees of freedom, yielding $C^\prime = 0.081(5)$. This seems to suggest the following combined scaling towards the continuum and large-$N$ limit of the autocorrelation time of $Q^2$ in the periodic system:
\beq\label{eq:tau0_scaling}
\tau_0(Q^2) \approx \left[ \frac{C^\prime}{\left(\elld\sqrt{\sigma}\right)^2}\right] \left(\frac{1}{a\sqrt{\sigma}}\right)^2 N^{0.5}, \qquad C^\prime = 0.081(5).
\eeq

\noindent This formula highlights the scaling of $\tau_0$ with $a$ and $N$ as the continuum limit and the large-$N$ limit are approached, and how the defect size influences the overall prefactor of $\tau_0$. To investigate this point further, I have also performed a combined best fit according to:
\beq
\tau_0(Q^2) = \left[ \frac{C^\prime}{\left(\elld\sqrt{\sigma}\right)^{A_1}}\right] \left(\frac{1}{a\sqrt{\sigma}}\right)^{A_2} \left(\frac{1}{N}\right)^{A_3},
\eeq
yielding, when fixing $C^\prime = 0.08$, $A_1 = 2.72(54)$, $A_2 = 1.9(2)$, $A_3 = 0.3(3)$, and a reduced chi-squared of $\sim 1.3$ with 8 degrees of freedom. These exponents agree with what I have empirically determined. In conclusion, the scaling of the figure of merit $\tau(Q^2) = \Nr \tau_0(Q^2)$ in the explored parameter space for $a$, $N$ and $\elld$ can be roughly approximated by:
\beq\label{eq:tau_scaling}
\tau(Q^2) = \Nr\tau_0(Q^2) \approx \left[ \frac{0.065}{\left(\elld\sqrt{\sigma}\right)^{0.5}}\right] \left(\frac{1}{a\sqrt{\sigma}}\right)^{3.5} N^{1.5},
\eeq
obtained combining Eq.~\eqref{eq:Nr_scaling} and Eq.~\eqref{eq:tau0_scaling}. In the explored range of defect sizes $\elld\sqrt{\sigma}\sim 0.45-0.55$, the dependence of $\tau(Q^2)=\Nr \tau_0(Q^2)$ on $\elld\sqrt{\sigma}$ is very mild and almost completely cancels between $\tau_0$ and $\Nr$, as already observed in previous studies~\cite{Bonanno:2020hht,Bonanno:2024nba,Bonanno:2024ggk,Bonanno:2024zyn}.

These results for the PTBC autocorrelation times can be compared with previous ones reported in the literature with periodic and open boundary conditions (PBC/OBC):
\begin{itemize}
\item 
In Ref.~\cite{Athenodorou:2021qvs}, the authors report that, for $N=6$ and $\beta=26.22$, the next-to-finest lattice spacing considered in this study, the average number of lattice sweeps needed to observe a $\pm 1$ jump of the lattice topological charge with PBC is $\sim 10^6$, which is respectively 4 and 3 orders of magnitude larger than the values of $\tau_0(Q^2)$ and $\tau(Q^2)$ I find for the same simulation point. Due to these huge autocorrelations, the authors of Ref.~\cite{Athenodorou:2021qvs} only report determinations of $\chi$ in the SU(6) theory up to $\beta=25.32$, which is the coarsest lattice spacing I have considered for $N=6$.
\item
In Ref.~\cite{Ce:2016awn}, the authors adopt OBC and report determinations of $\chi$ for SU(6) up to $\beta=26.15$, a bit coarser than my next-to-finest point. They do not report a result for the autocorrelation time of $Q^2$, but observe that separating subsequent measures of $Q^2$ with 450 updating steps at this $\beta$ value are sufficient to observe small autocorrelations, thus this number can be taken as an estimate of $\tau(Q^2)$. In this case they use 12 over-relaxation sweeps per each heat-bath one, and a lattice where the temporal extent is 4 times the spatial one to avoid boundary effects. Thus, their updating step is 12 times more costly than the one I've used, leading to $\tau(Q^2) \sim 5400$ in the units adopted here. This is $\sim 4$ times larger than the value $\tau(Q^2) = \Nr \tau_0(Q^2) \sim 1270$ I obtain interpolating my data at $(a\sqrt{\sigma})(\beta=26.15)$ using Eq.~\eqref{eq:tau_scaling}.
\end{itemize}
\newpage

\begin{figure}[!t]
\centering
\includegraphics[scale=0.295]{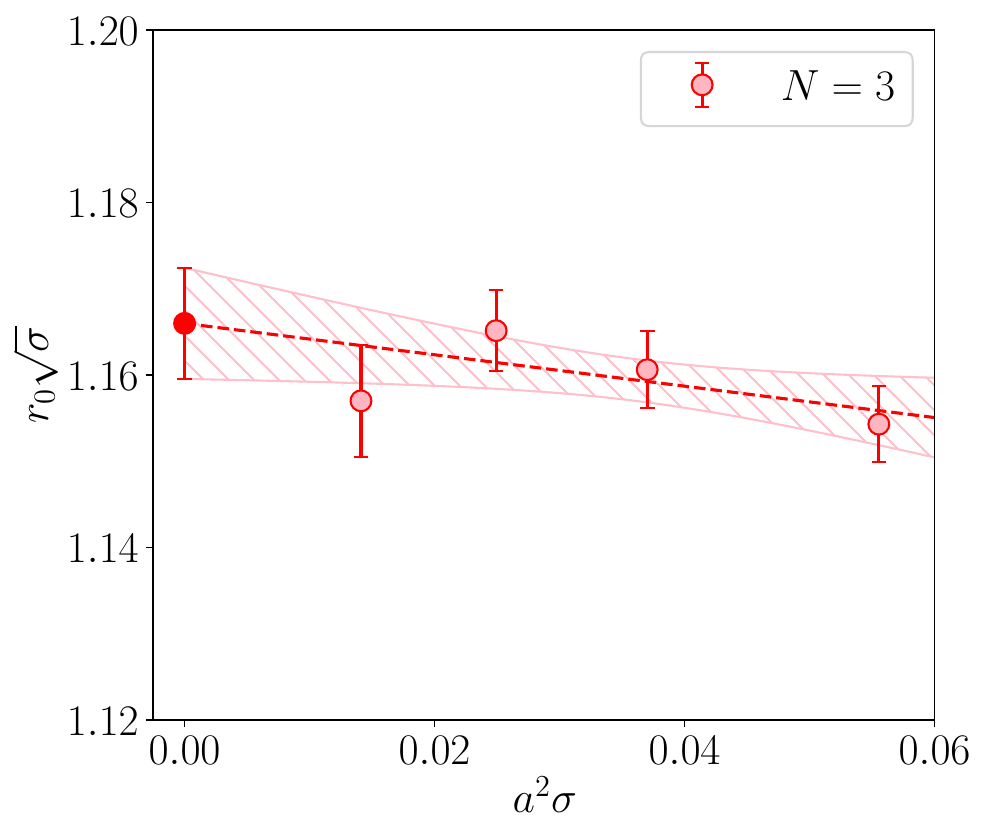}
\includegraphics[scale=0.295]{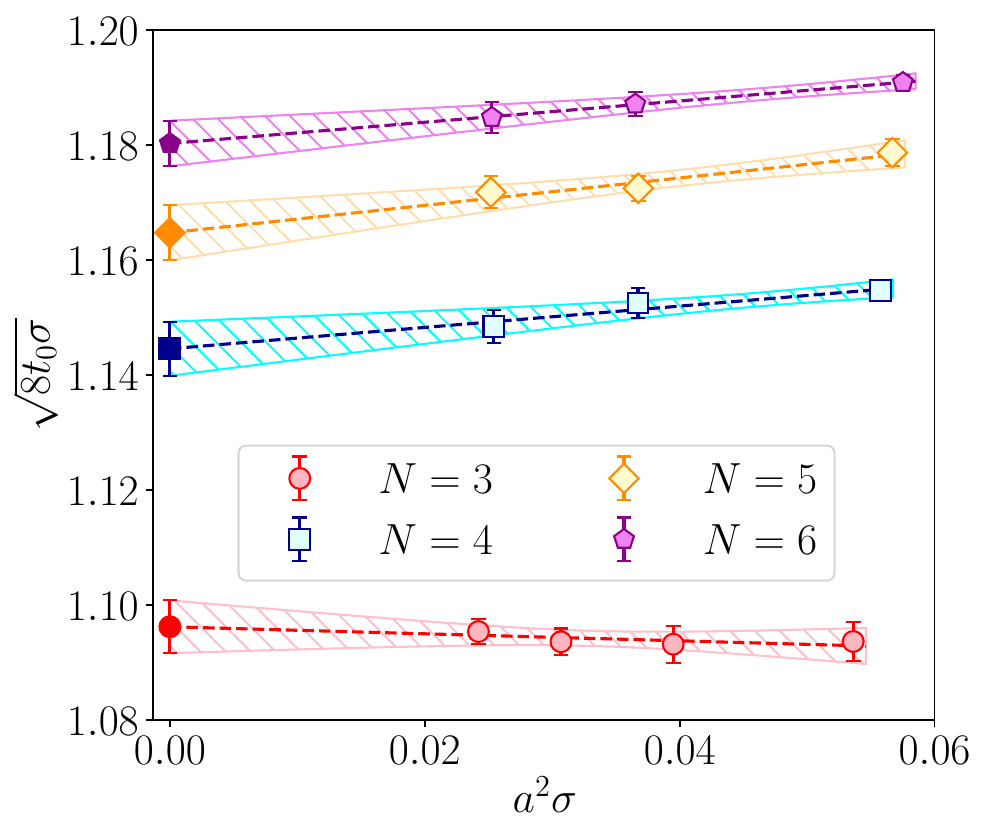}
\includegraphics[scale=0.295]{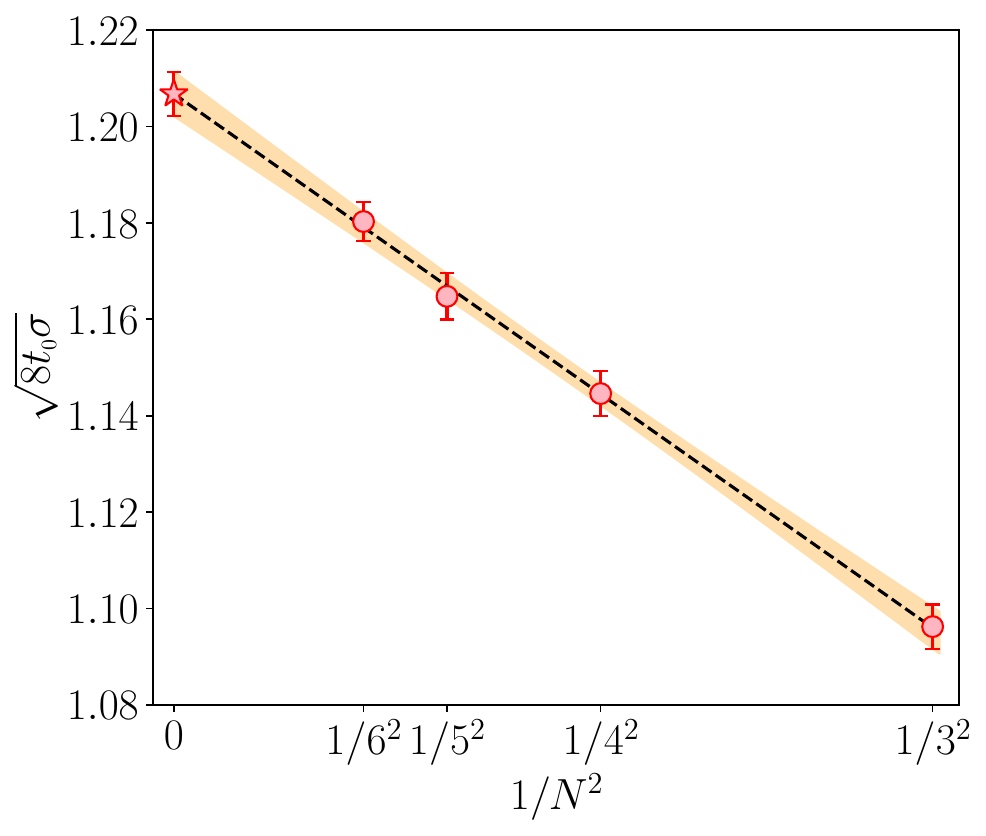}
\caption{Computation of the continuum limit of $r_0\sqrt{\sigma}$ for $N=3$ (left panel), and of $\sqrt{8t_0\sigma}$ for $N=3,4,5,6$ (center panel) and $N=\infty$ (right panel).}
\label{fig:scale_conv}
\end{figure}

\begin{table}[!t]
\begin{center}
\begin{tabular}{|c|c|c|}
\hline
&&\\[-1em]
Ratio of scales & $N$ & Value \\
\hline
\hline
$r_0\sqrt{\sigma}$ & 3 & 1.1660(64)\\
\hline
\hline
\multirow{5}{*}{$\sqrt{8t_0\sigma}$} & 3 & 1.0962(46) \\
& 4 & 1.1446(47) \\
& 5 & 1.1648(48) \\
& 6 & 1.1803(40) \\
& $\infty$ & 1.2068(46) \\
\hline
\end{tabular}
\end{center}
\caption{Continuum-extrapolated conversion factors between the Sommer scale $r_0$, the gradient flow scale $\sqrt{8t_0}$ and the string tension $\sqrt{\sigma}$. At finite lattice spacing the ratios are  obtained interpolating results for $\sqrt{\sigma}$ of~\cite{Athenodorou:2020ani,Athenodorou:2021qvs} at the $\beta$ values of~\cite{Necco:2001xg} ($r_0$) and of~\cite{Ce:2015qha,Ce:2016awn,Giusti:2018cmp} ($\sqrt{8t_0}$), see Fig.~\ref{fig:scale_conv}.}
\label{tab:scale_conv}
\end{table}

Finally, before moving on with the presentation of the PTBC results for the topological susceptibility, given that in the following I will compare them with prior ones found in the literature, I will conclude this section by discussing the computation of a few scale setting conversion factors. These will be helpful to translate previous determinations of $\chi$ in terms of $\sigma$. In particular, I have considered the Sommer parameter $r_0$~\cite{Guagnelli:1998ud,Necco:2001xg,Sommer:2014mea} and the gradient flow reference scale $\sqrt{8t_0}$~\cite{Luscher:2010iy}. Continuum-extrapolated results for the ratio of these scales with $\sqrt{\sigma}$, obtained performing a linear fit according to leading $\mathcal{O}(a^2)$ lattice artifacts for $a^2\sigma < 0.06$, can be found in Tab.~\ref{tab:scale_conv}, and are shown in Fig.~\ref{fig:scale_conv}. Results at finite lattice spacing are obtained interpolating the string tension determinations of Refs.~\cite{Athenodorou:2020ani,Athenodorou:2021qvs} at the $\beta$ values of Ref.~\cite{Necco:2001xg} (for $r_0$ at $N=3$), of Ref.~\cite{Ce:2015qha} (for $\sqrt{8t_0}$ at $N=3$), and of Ref.~\cite{Ce:2016awn} (for $\sqrt{8t_0}$ at $N=4,5,6$). As a consistency check at $N=3$, I observe that $\sqrt{8t_0}/r_0 = \sqrt{8t_0\sigma}/r_0\sqrt{\sigma} = 0.9401(65)$  agrees well with the previous results $\sqrt{8t_0}/r_0=0.941(7)$~\cite{Ce:2015qha} and $\sqrt{8t_0}/r_0=0.950(10)$~\cite{Giusti:2018cmp}. Finally, for $\sqrt{8t_0\sigma}$, given that this ratio of scale is available for several values of the number of colors, I am also able to obtain a result in the $N=\infty$ limit from a large-$N$ extrapolation, assuming leading $1/N^2$ corrections to $N=\infty$. Such calculation yields, cf.~Fig.~\ref{fig:scale_conv}:
\beq
\sqrt{8t_0\sigma}(N) = 1.2068(46) - \frac{0.997(69)}{N^2} + \mathcal{O}\left(\frac{1}{N^4}\right).
\eeq
\FloatBarrier
\newpage
\clearpage
\subsection{Determination of $\chi$ at large $N$}

The goal of this section is to discuss the continuum and large-$N$ limits of $\chi$, and to verify the independence of $\chi$ on the choice of the smoothing radius in the continuum limit.

First of all, I have computed the topological susceptibility according to the discretization discussed in Sec.~\ref{sec:setup_topsusc} for several values of the smoothing radius in the range $0.85 \lesssim \Rs\sqrt{\sigma} \lesssim 1.6$. The lower bound of the range is chosen in order to ensure that $n_\cool > 5$ for the smallest lattice spacing employed, which translates into a smoothing radius which is at least 4 lattice spacings for all simulation points. This seems a reasonable choice to avoid UV contamination in the related lattice operator. The upper bound is instead dictated by the maximum distance $\frac{1}{2}\ell\sqrt{\sigma} \sim 1.7-1.85$ that can be reached on a periodic box of size $\ell\sqrt{\sigma}\sim 3.4-3.7$. All lattice determinations of $\chi_{\L}(\Rs)$ as a function of $\beta$ and $N$ can be found in Appendix~\ref{app:rawdata}.

\begin{figure}[!t]
\centering
\includegraphics[scale=0.305]{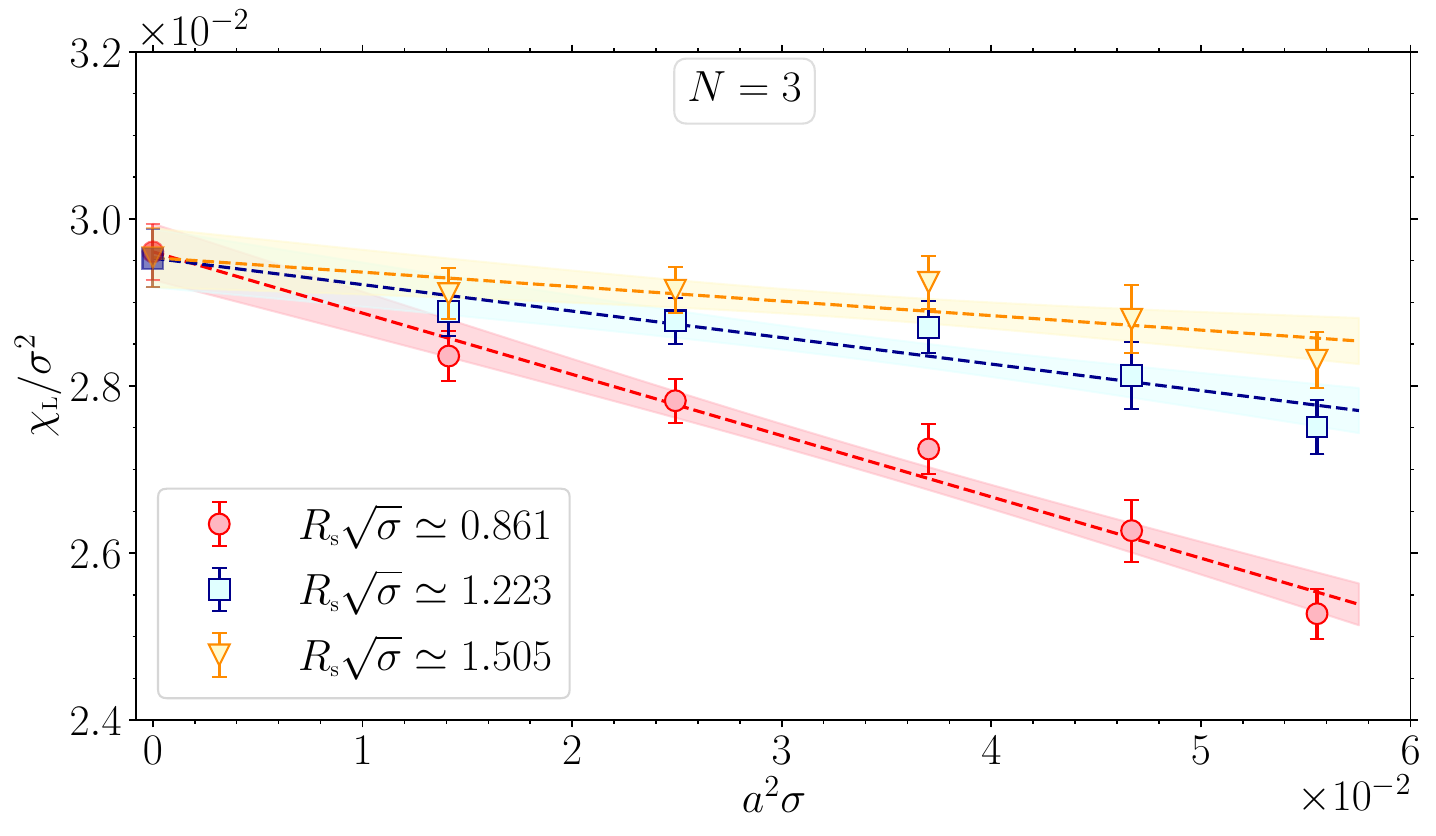}
\includegraphics[scale=0.305]{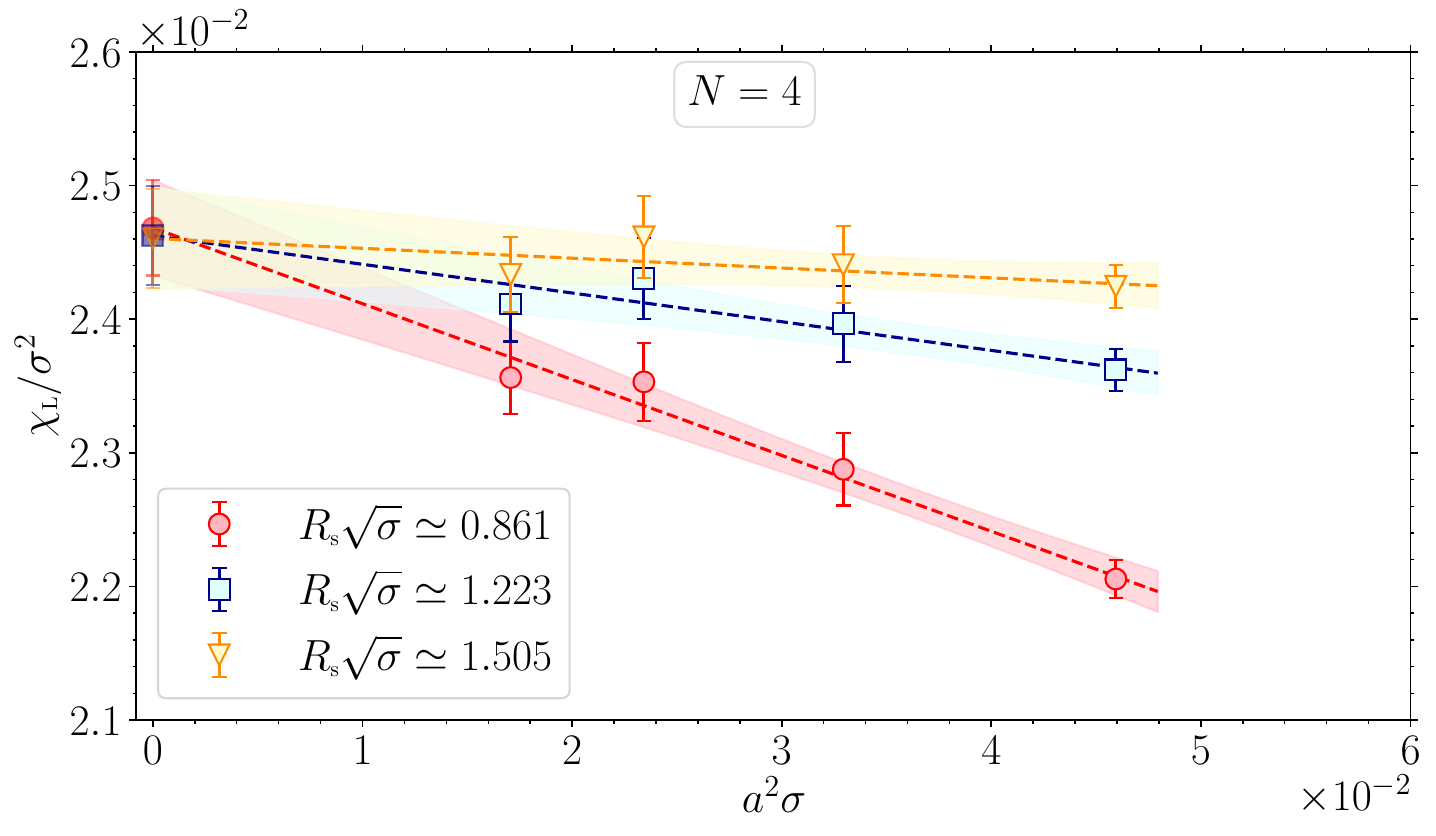}
\includegraphics[scale=0.305]{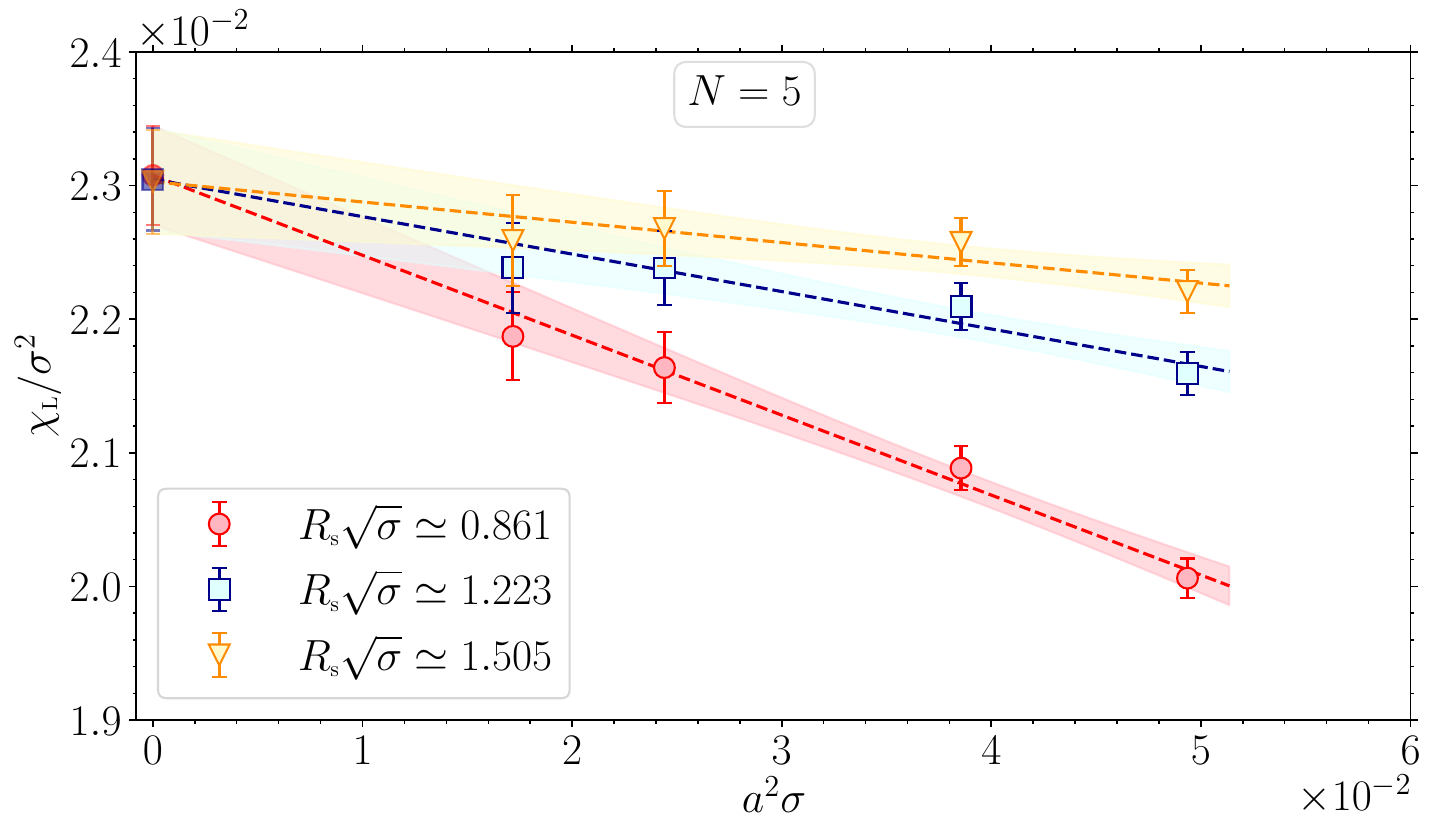}
\includegraphics[scale=0.305]{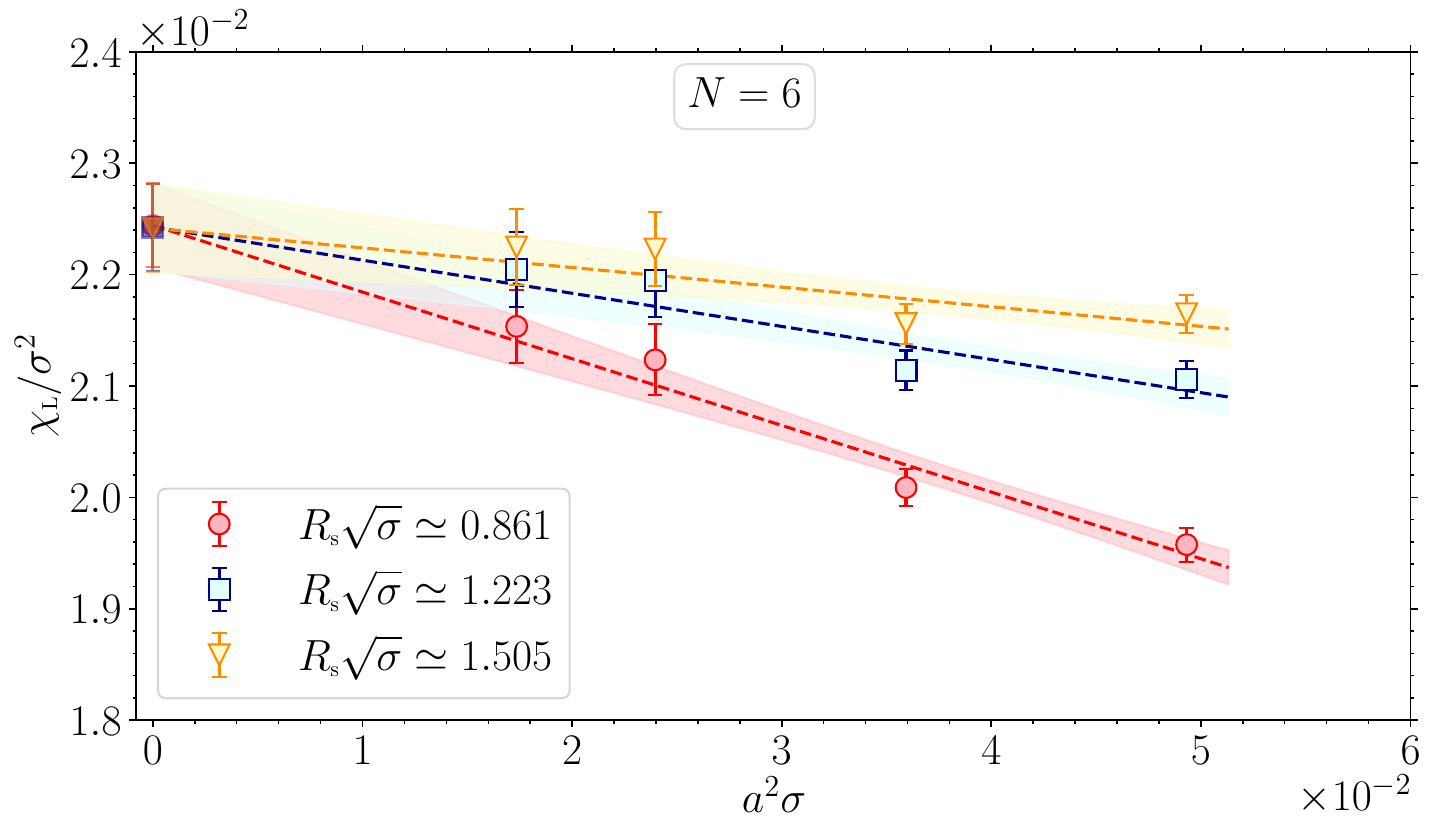}
\caption{Continuum limits of $\chi_{\L}$ for various values of $N$ for 3 values of the smoothing radius $\Rs$. This quantity is kept fixed as a function of both $a$ and $N$. Each continuum limit is taken according to the fit function in Eq.~\eqref{eq:fit_function_contlim_N_by_N}.}
\label{fig:contlim_chi_N_by_N_1}
\end{figure}

I will start my discussion by performing individual analyses of each ensemble at fixed $N$. For any given number of colors, I will take the continuum limit of $\chi_{\L}/\sigma^2$ at fixed value of $\Rs\sqrt{\sigma}$ for a few choices of the smoothing radius in order to check the expected continuum scaling:
\beq\label{eq:fit_function_contlim_N_by_N}
\frac{1}{\sigma^2}\chi_{\L}(a,N,\Rs) = \frac{1}{\sigma^2}\chi(N) + C(\Rs,N) a^2 \sigma + \dots.
\eeq
To do so, I have interpolated the data for $\chi_{\L}$ in Appendix~\ref{app:rawdata} as a function of $\Rs$ using a cubic spline interpolation, and considered determinations at different values of $\beta$ and $N$ possessing the same value of $\Rs\sqrt{\sigma}$ up to the third digit, which is where the error on the smoothing radius sits given the $\sim \mathcal{O}(0.1-0.2\%)$ precision on the adopted scale setting. As it can be seen from Fig.~\ref{fig:contlim_chi_N_by_N_1}, my data can be perfectly described by a linear $\mathcal{O}(a^2)$ behavior across the whole range of lattice spacings and smoothing radii explored. At finite lattice spacing, the lattice topological susceptibility does depend on $\Rs$, but this dependence becomes weaker and weaker as the continuum limit is approached. Thus, while the slope of lattice artifacts $C(\Rs,N)$ depends sizeably on the choice of $\Rs$, I observe perfect independence of the continuum extrapolated results $\chi(N)$ on the smoothing radius, as expected.

\begin{figure}[!t]
\centering
\includegraphics[scale=0.305]{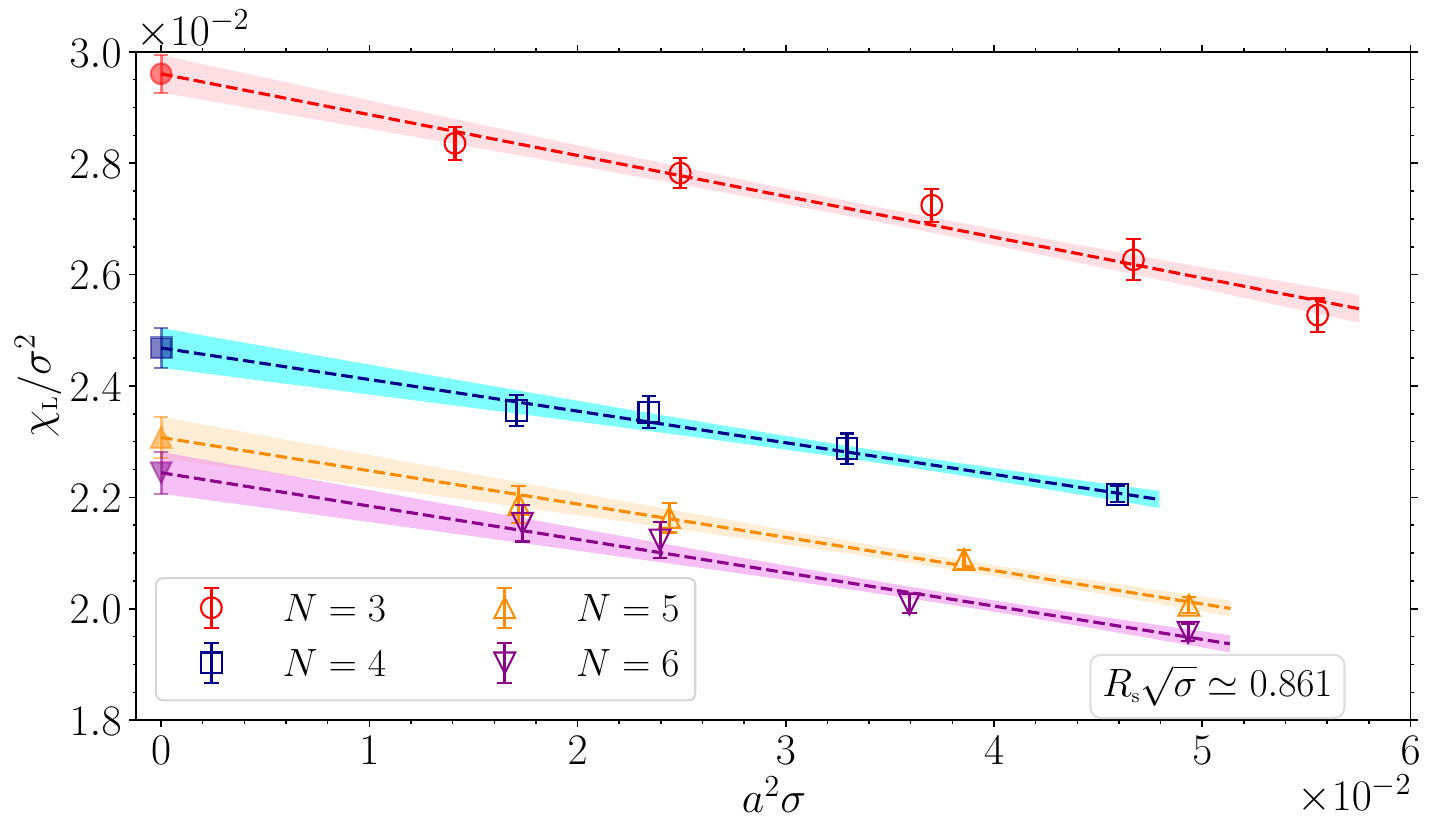}
\includegraphics[scale=0.305]{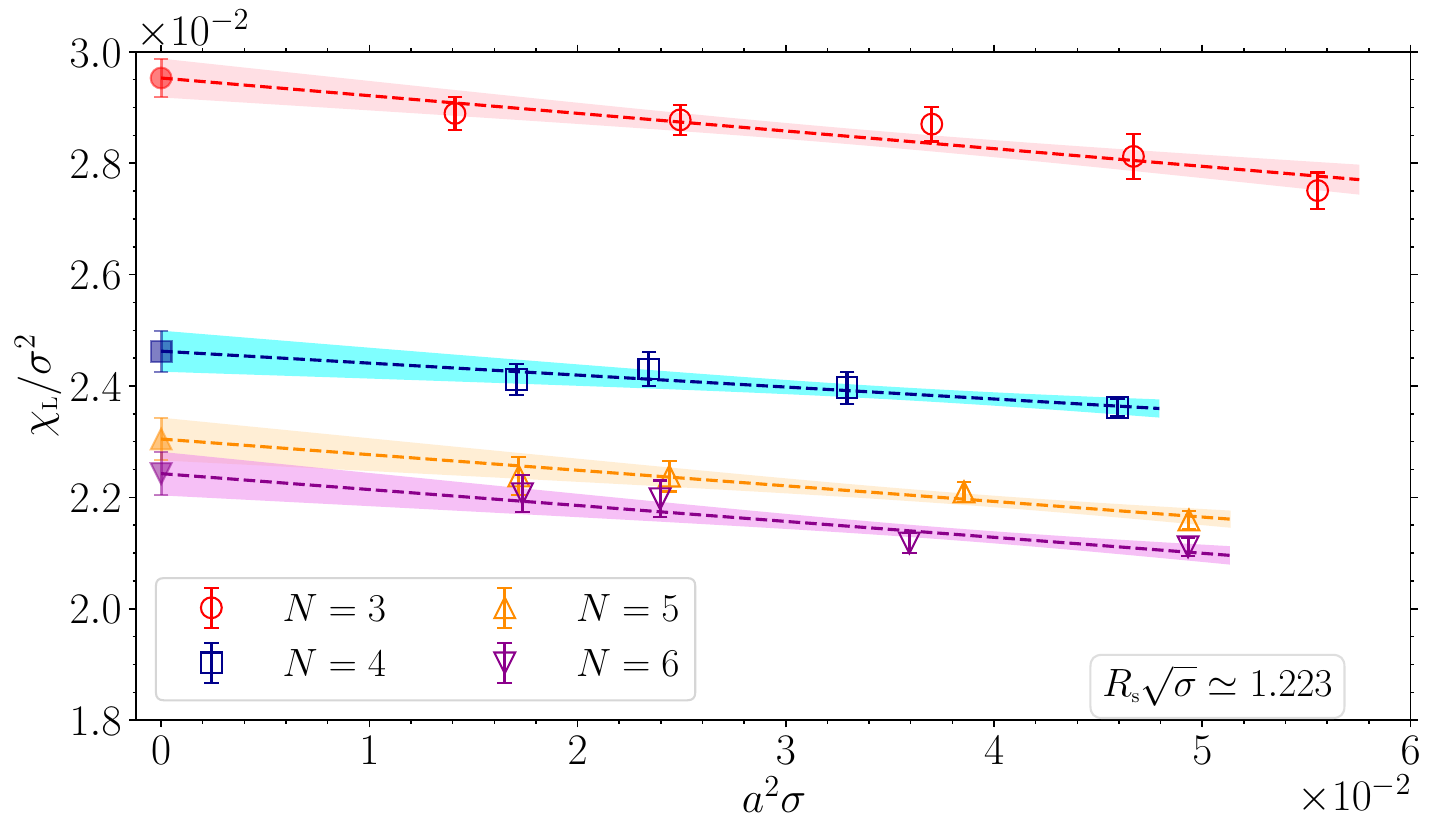}
\includegraphics[scale=0.305]{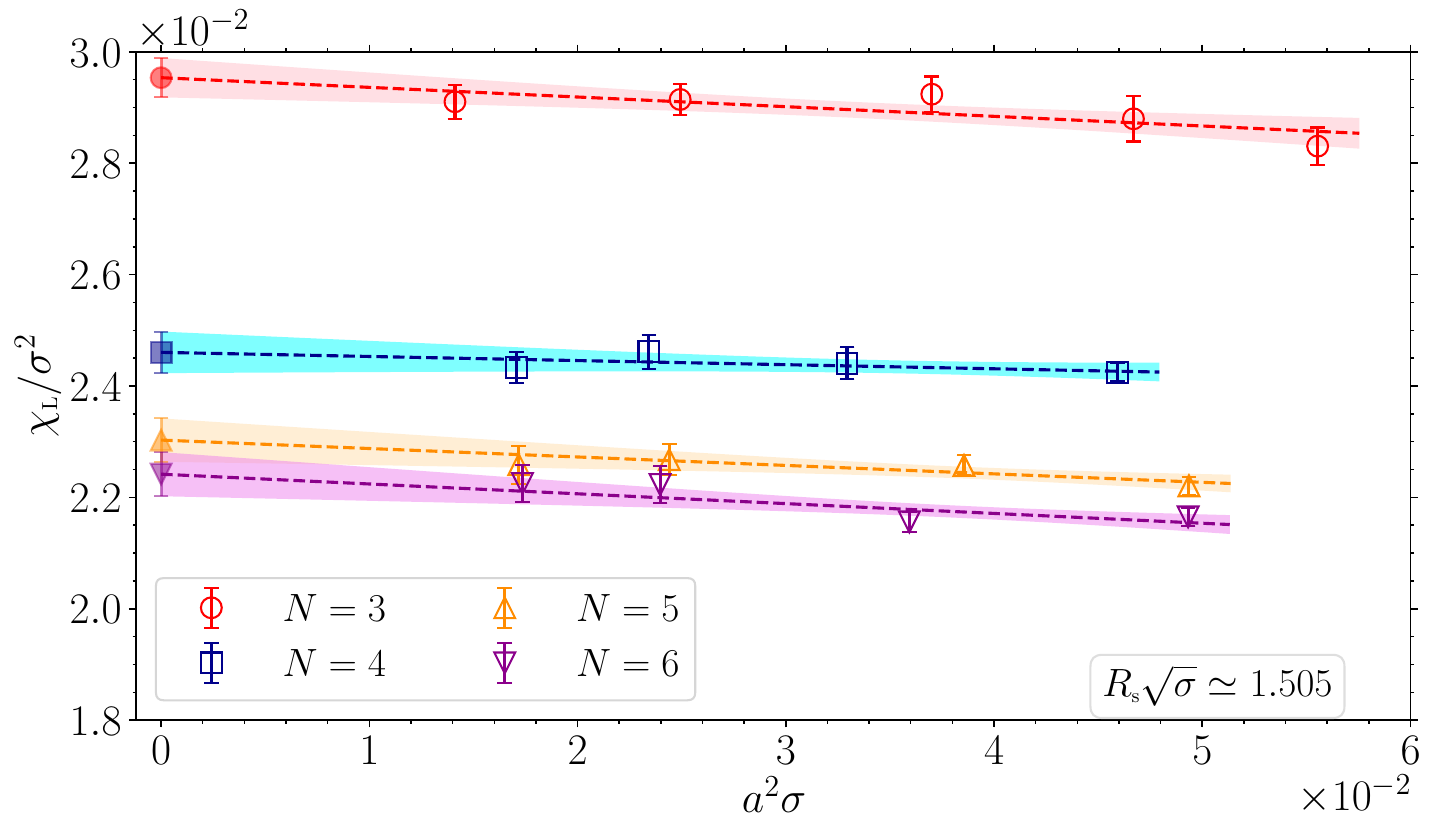}
\includegraphics[scale=0.305]{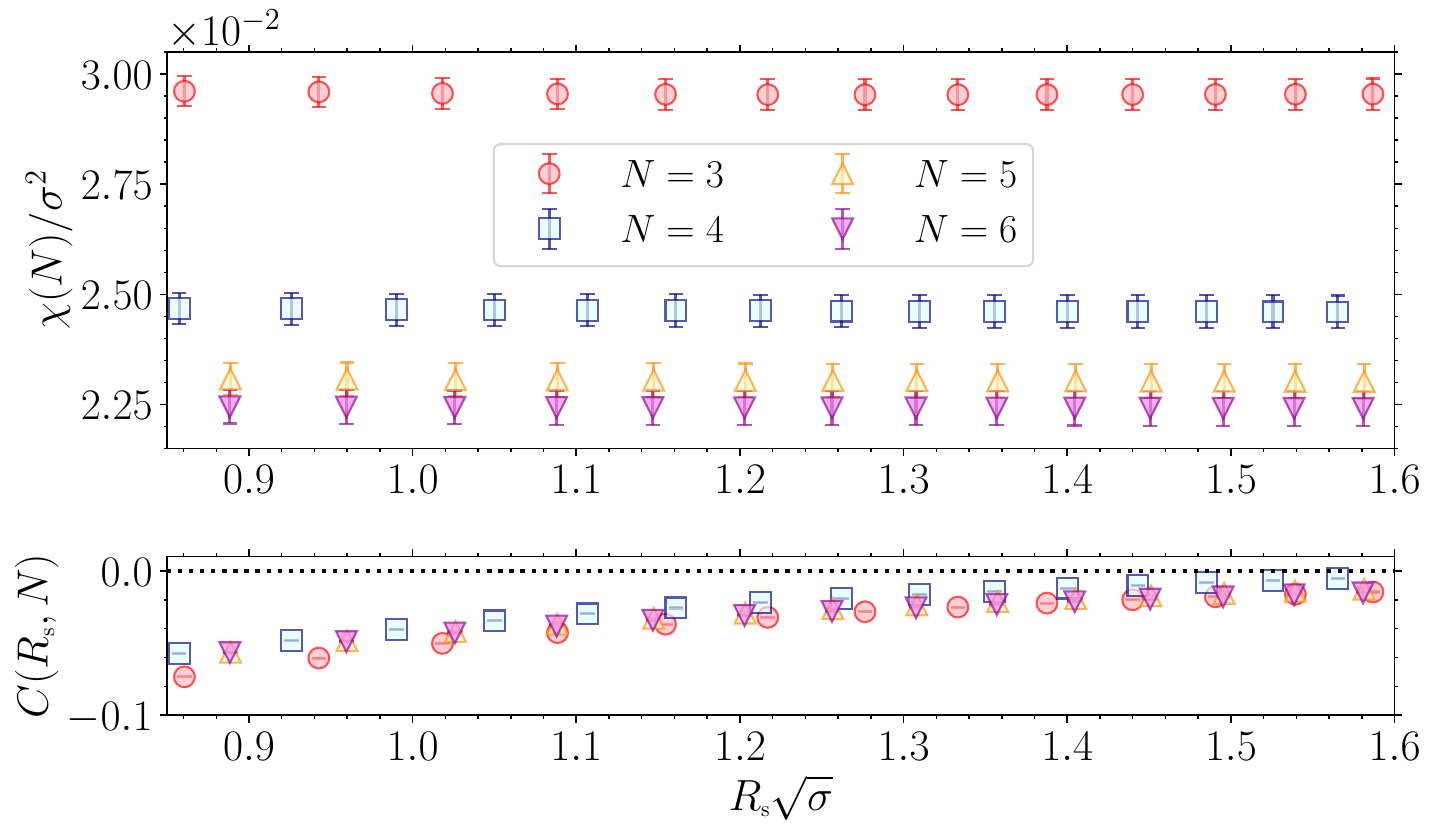}
\caption{Same data as in Fig.~\ref{fig:contlim_chi_N_by_N_1}, but plotted for various values of $N$ at fixed $\Rs\sqrt{\sigma}$. Lattice artifact slopes $C(\Rs,N)$ are practically independent of $N$, and become very small for $\Rs\sqrt{\sigma}\gtrsim 1.2$. In the bottom right plot I also show the full $\Rs$-dependence of $\chi(N)/\sigma^2$ and $C(\Rs,N)$ in the whole explored range of smoothing radii.}
\label{fig:contlim_chi_N_by_N_2}
\end{figure}

Interestingly, lattice artifacts seem to be practically independent of $N$ for all explored values of $\Rs$---i.e., $C(\Rs,N)\simeq C(\Rs)$---as revealed by the plots in Fig.~\ref{fig:contlim_chi_N_by_N_2}, where I compare the continuum limits of $\chi_\L$ as a function of $N$ at fixed smoothing radius. This fact was already observed in~\cite{Ce:2016awn}. Moreover, lattice artifacts become very small above $\Rs\sqrt{\sigma}\gtrsim 1.2$, as it can be seen from Figs.~\ref{fig:contlim_chi_N_by_N_1} and~\ref{fig:contlim_chi_N_by_N_2}. For example, for the value $\Rs\sqrt{\sigma}\simeq 1.223$ reported in those figures, the coarsest lattice spacing determinations of $\chi_{\L}$ differ from their corresponding continuum ones by $\sim 2\%$ at most for all values of $N$. Since the lattice artifacts affecting $\chi_{\L}(\Rs)$ have 3 different sources (the discretized lattice action used in the Monte Carlo, the discretized lattice action used in the smoothing algorithm, and the discretized observable used to compute the topological charge), the flattening of $C(\Rs)$ for large enough smoothing radii could be due to an accidental cancellation among them.\footnote{See Ref.~\cite{Ramos:2015baa} for a similar discussion about the lattice artifacts affecting the energy density computed after the gradient flow within the framework of Symanzik improvement.}

\FloatBarrier

\begin{table}[!t]
	\small
\begin{center}
\begin{tabular}{|c|c|c|c|c|}
\hline
$N$ & $\beta$ & $a\sqrt{\sigma}$ & $a^4 \chi_{\L} \times 10^5$ & $\chi_{\L}/\sigma^2$ \\
\hline
\hline
\multirow{5}{*}{3} &  5.95 &  0.23567(69)  &  8.486(12)  &  0.02751(32) \\
&  6.00   &  0.21609(76)  &  6.132(10)  &  0.02812(40) \\
&  6.07   &  0.19238(51)  &  3.9315(90)  &  0.02870(31) \\
&  6.20   &  0.15788(31)  &  1.7881(94)  &  0.02878(27) \\
&  6.40   &  0.11879(26)  &  0.5754(32)  &  0.02890(30) \\
\hline
\hline
\multirow{4}{*}{4}  &  11.02  &  0.21434(28)  &  4.985(20)  &  0.02362(16) \\
&  11.20  &  0.18149(49)  &  2.600(13)  &  0.02397(28) \\
&  11.40  &  0.15305(34)  &  1.334(12)  &  0.02430(30) \\
&  11.60  &  0.13065(21)  &  0.7026(68)  &  0.02411(28) \\
\hline
\hline
\multirow{4}{*}{5} &  17.43  &  0.22217(37)  &  5.261(16)  &  0.02159(16) \\
&  17.63  &  0.19636(35)  &  3.285(12)  &  0.02210(18) \\
&  18.04  &  0.15622(38)  &  1.333(10)  &  0.02238(28) \\
&  18.38  &  0.13106(30)  &  0.6566(89)  &  0.02238(33) \\
\hline
\hline
\multirow{4}{*}{6} &  25.32  &  0.22208(35)  &  5.135(24)  &  0.02111(16) \\
&  25.70  &  0.18956(33)  &  2.735(13)  &  0.02118(18) \\
&  26.22  &  0.15480(36)  &  1.262(15)  &  0.02197(33) \\
&  26.65  &  0.13173(29)  &  0.6656(84)  &  0.02206(34) \\
\hline
\end{tabular}
\end{center}
\caption{Results for $\chi_{\L}$ at $\Rs\sqrt{\sigma}\simeq 1.223$ both in lattice and physical units.}
\label{tab:res_chi_fix_Rs}
\end{table}

\begin{table}[!t]
\begin{center}
\begin{tabular}{|c|c|}
\hline
$N$ & $\chi/\sigma^2$ [$N$-by-$N$ fits] \\
\hline
3        & 0.02953(35)  \\
4        & 0.02463(37)  \\
5        & 0.02305(38)  \\
6        & 0.02242(39)  \\
\hline
\end{tabular}
\end{center}
\caption{Continuum limit results for $\chi/\sigma^2$ as a function of $N$ obtained from the individual $N$-by-$N$ fits for $\Rs\sqrt{\sigma} \simeq 1.223$.}
\label{tab:chi_cont_individual}
\end{table}


In light of these results, from now on I will just choose one value of the smoothing radius and keep it fixed to quote my results, as I have shown clear evidence that this is irrelevant in the continuum limit. To this end, I will choose $\Rs\sqrt{\sigma}\simeq 1.223 \simeq 0.5$ fm, which lies in the middle of the range of smoothing radii I have explored, and for which lattice artifacts are pretty mild. For completeness, I have reported all finite-lattice-spacing determinations for this choice of the smoothing radius in Tab.~\ref{tab:res_chi_fix_Rs}, and their corresponding continuum extrapolations in Tab.~\ref{tab:chi_cont_individual}.

\begin{figure}[!t]
\centering
\includegraphics[scale=0.55]{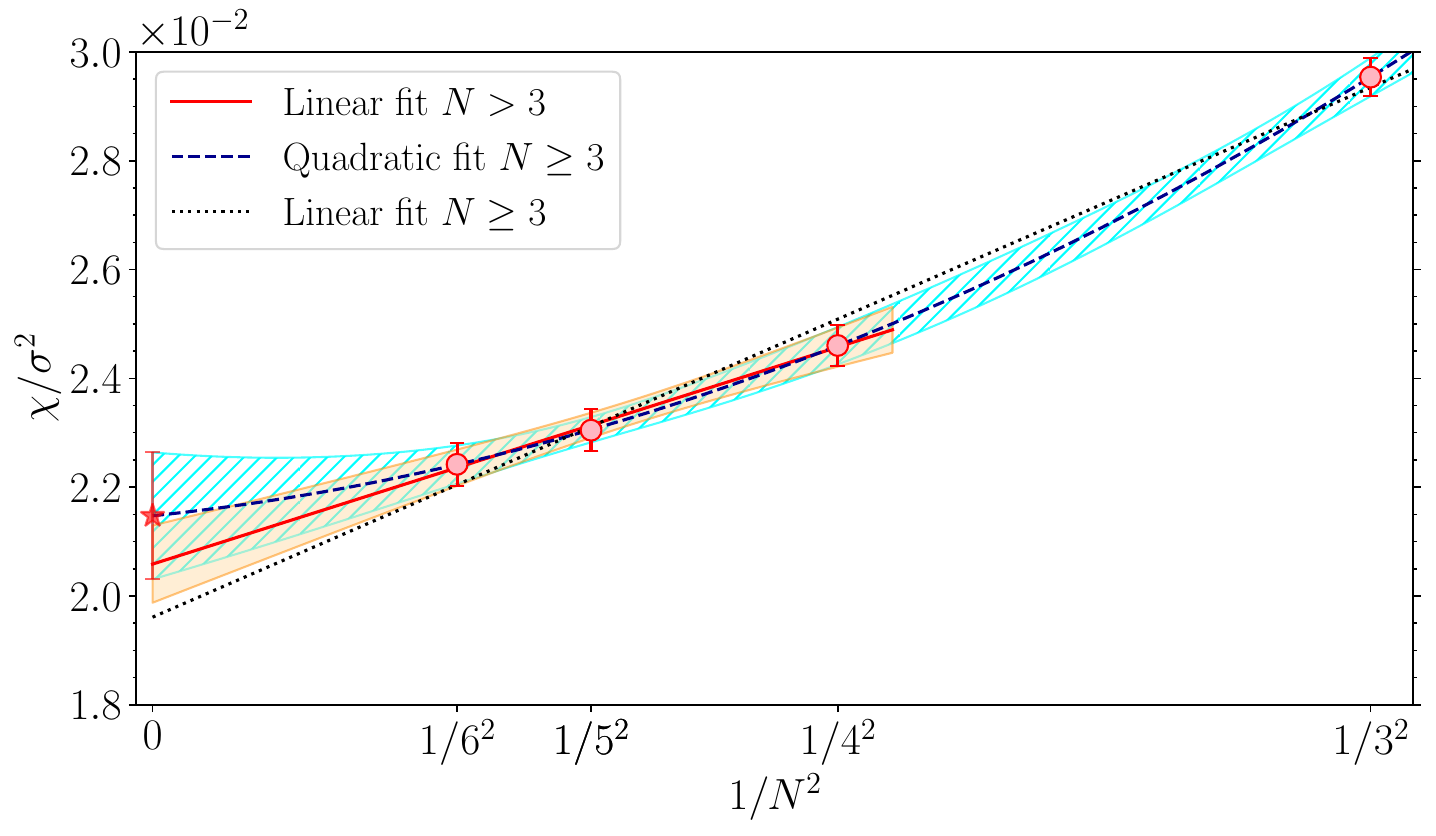}
\caption{Large-$N$ extrapolation of the continuum determinations of $\chi$ obtained from the individual $N$-by-$N$ fits in Fig.~\ref{fig:contlim_chi_N_by_N_1}. I reported the results of linear and quadratic fits in $1/N^2$ in several fit ranges, and the final $N=\infty$ determination in $1/N=0$ (see the text for more details).}
\label{fig:largeN_individual_MYRES}
\end{figure}

The obtained continuum results for $\chi(N)/\sigma^2$ can be now extrapolated towards the large-$N$ limit, assuming an expansion of the topological susceptibility in powers of $1/N^2$. To this end I will consider the following fit function:
\beq\label{eq:fit_function_chi_vs_N}
\frac{1}{\sigma^2}\chi(N) = \frac{1}{\sigma^2}\chi_{\inf} + \frac{A_2}{N^2} + \frac{A_4}{N^4} + \mathcal{O}\left(\frac{1}{N^6}\right).
\eeq
As it can be seen from Fig.~\ref{fig:largeN_individual_MYRES}, the collected data can be very well described by Eq.~\eqref{eq:fit_function_chi_vs_N}. The best fit of Eq.~\eqref{eq:fit_function_chi_vs_N} to the data yields:
\beq
\frac{1}{\sigma^2}\chi(N) = 0.0215(12) + \frac{0.021(39)}{N^2} + \frac{0.46(27)}{N^4} + \mathcal{O}\left(\frac{1}{N^6}\right), \qquad (N\ge 3).
\eeq
Removing the $N=3$ data from the fitted data and the $1/N^4$ term from the fit function~\eqref{eq:fit_function_chi_vs_N}, perfectly compatible results are obtained:
\beq
\frac{1}{\sigma^2}\chi(N) = 0.02059(71) + \frac{0.064(15)}{N^2} + \mathcal{O}\left(\frac{1}{N^4}\right), \qquad \qquad (N>3).
\eeq
Actually, one could in principle also fit the whole data set assuming just a linear $1/N^2$ correction to the large-$N$ limit. This would yield:
\beq
\frac{1}{\sigma^2}\chi(N) = 0.01961(40) + \frac{0.088(6)}{N^2} + \mathcal{O}\left(\frac{1}{N^4}\right), \qquad \qquad (N\ge3),
\eeq
but a much larger chi-squared compared to the quadratic fit earlier discussed. Moreover, the coefficient of the $1/N^2$ correction turns out to be somewhat larger compared to the one obtained from the linear best fit in the restricted range $N>3$, leading to a somewhat lower $N=\infty$ extrapolated value $\chi_{\inf}$. Although the latter is not incompatible within errors with the extrapolations previously obtained, the collected evidence seems to hint at the presence of higher-order terms affecting the $N=3$ determination, and at the fact that $N=3$ falls out of the regime where finite-$N$ data can be described with a simple $1/N^2$ term. Therefore, the result of this best fit will not be considered further.

In the end, in order to be conservative, for the $N$-by-$N$ analysis I will take the result obtained from the quadratic fit performed in the whole range as the large-$N$ limit of the topological susceptibility:
\beq
\frac{1}{\sigma^2}\chi_{\inf} = 0.0215(12),
\eeq
since it has the largest statistical error. This result is also reported in Fig.~\ref{fig:largeN_individual_MYRES} as a star point in $1/N=0$.

\begin{figure}[!t]
\centering
\includegraphics[scale=0.5]{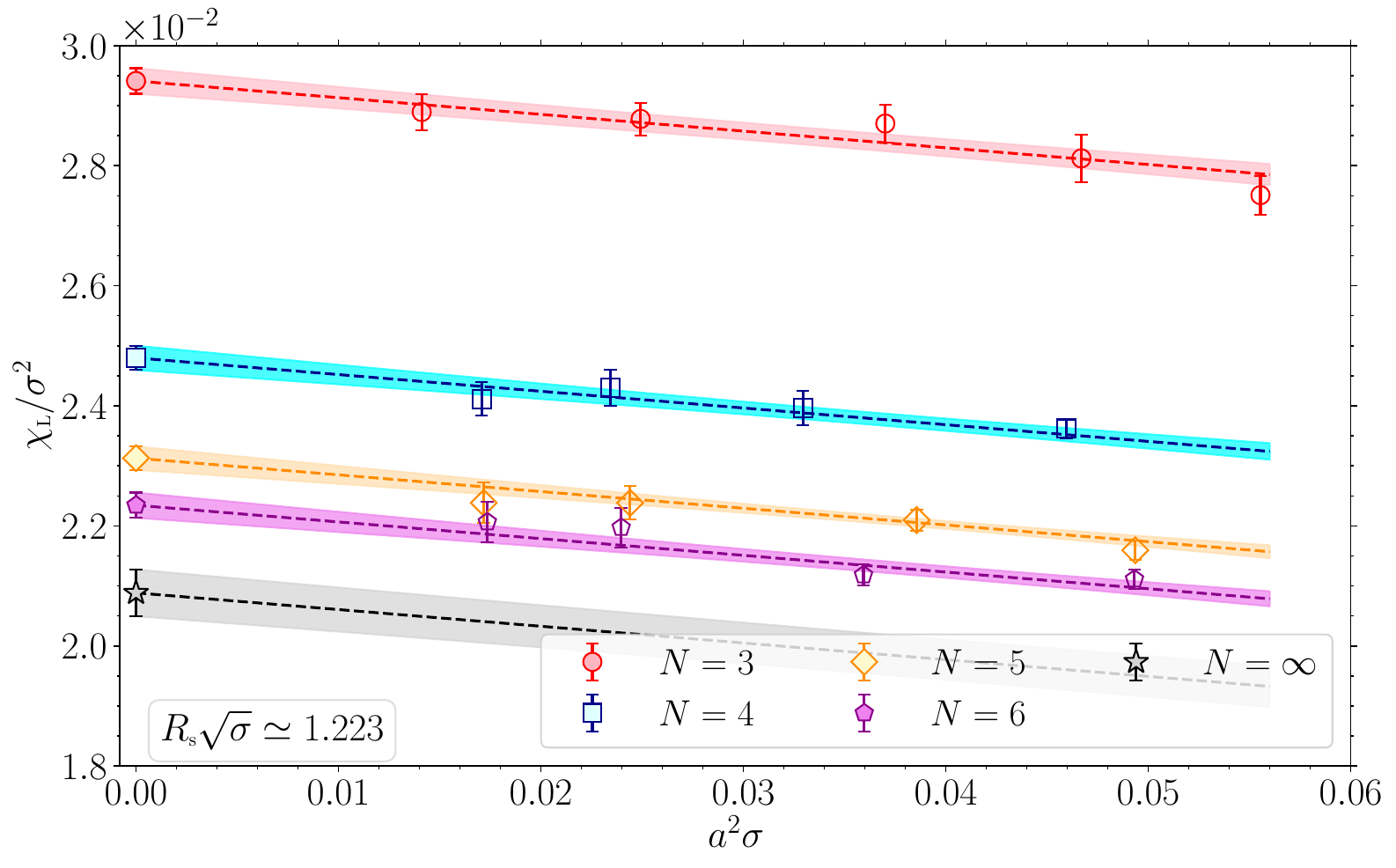}
\caption{Global best fit to achieve the simultaneous continuum-large-$N$ extrapolations of the data in Tab.~\ref{tab:res_chi_fix_Rs} ($\Rs\sqrt{\sigma}\simeq1.223$) according to the fit function in Eq.~\eqref{eq:fit_ansatz_GLOBAL}, which takes advantage of the observed $N$-independence of lattice artifacts.}
\label{fig:largeN_GLOBAL}
\end{figure}

\begin{figure}[!t]
\centering
\includegraphics[scale=0.5]{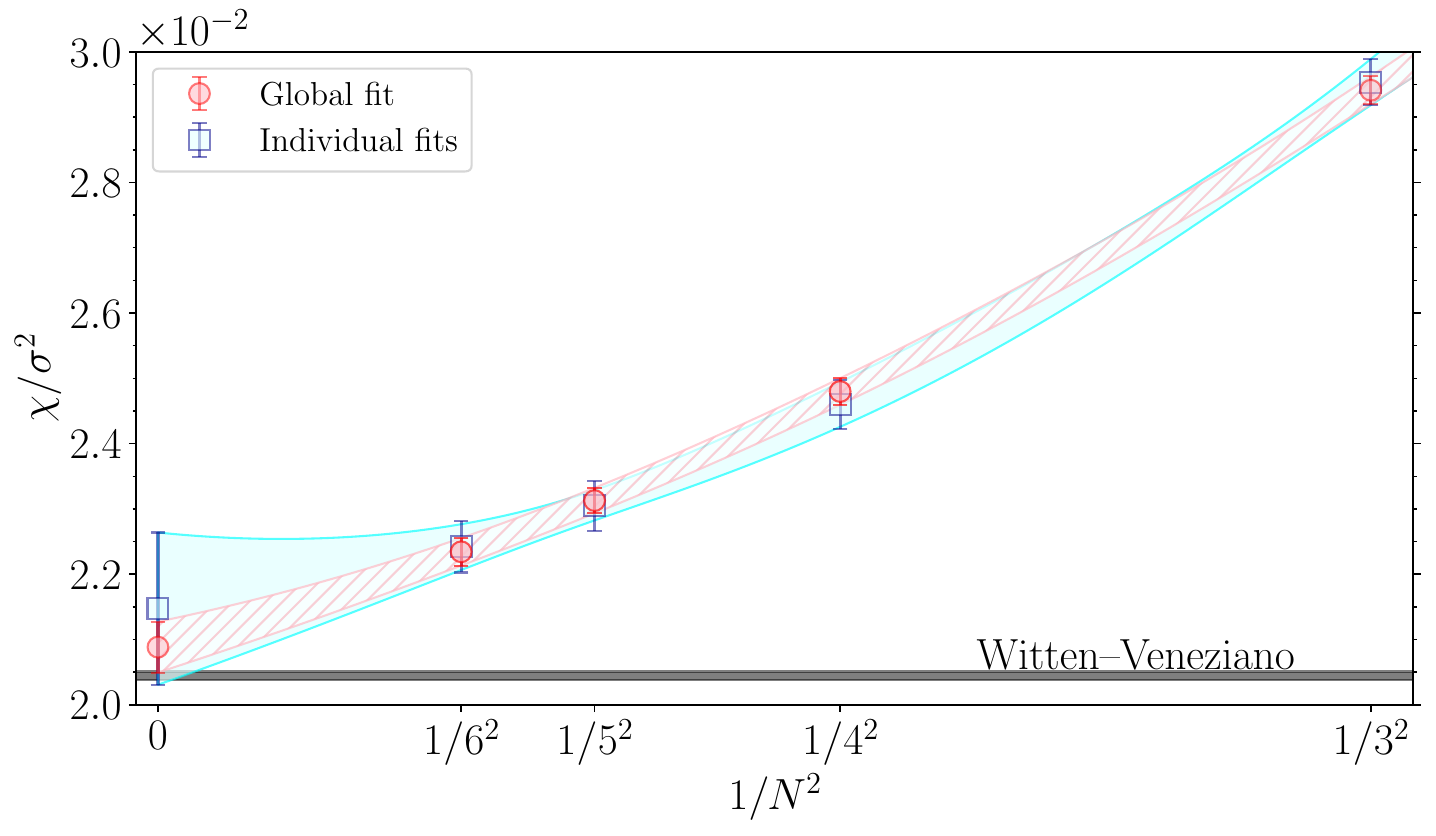}
\caption{Comparison among the continuum results for $\chi(N)$ obtained from the individual $N$-by-$N$ extrapolations and the global continuum-large-$N$ limit. The horizontal band is the traditional value $\chi_{\inf}\simeq\left(180~\mathrm{MeV}\right)^4$ obtained from the Witten--Veneziano formula~\eqref{eq:witten_veneziano_formula} and from $\sigma$ of~\eqref{eq:sigma_largeN}, see the text of Sec.~\ref{sec:res_discuss}.}
\label{fig:comp_chi_MYRES}
\end{figure}

Inspired by the discussion in Ref.~\cite{Ce:2016awn}, I have also performed a global analysis of my determinations of the topological susceptibility, where all ensembles at different values of $N$ are considered at the same time. In particular, following the lines of~\cite{Ce:2016awn}, and since the results shown in Fig.~\ref{fig:contlim_chi_N_by_N_2} point out the independence of lattice artifacts on $N$ regardless of the choice of the smoothing radius, I have performed a global continuum-large-$N$ extrapolation of the data for $\chi_{\L}(a,N)/\sigma^2$ in Tab.~\ref{tab:res_chi_fix_Rs} according to the following fit function:
\beq\label{eq:fit_ansatz_GLOBAL}
\frac{1}{\sigma^2}\chi_{\L}(a,N) = \frac{1}{\sigma^2}\chi_{\inf} + \frac{A_2}{N^2}+\frac{A_4}{N^4} + C a^2 \sigma.
\eeq
This ansatz provides an excellent description of the data with a chi-squared of 0.64 with 13 degrees of freedom ($p$-value $\simeq 82\%$), and the resulting best fit is shown in Fig.~\ref{fig:largeN_GLOBAL}. The global best fit yields for the $1/N$ expansion of $\chi(N)$:
\beq
\frac{1}{\sigma^2}\chi(N) = 0.02088(39) + 0.044(12) \frac{1}{N^2} + 0.293(83)\frac{1}{N^4} + \mathcal{O}\left(\frac{1}{N^6}\right).
\eeq

\begin{table}[!t]
\begin{center}
\begin{tabular}{|c|c|c|}
\hline
$N$ & $\chi/\sigma^2$ [Global fit] & $(8t_0)^2\chi$\\
\hline
3        & 0.02941(21) & 0.04248(77) \\
4        & 0.02479(20) & 0.04256(78) \\
5        & 0.02312(20) & 0.04257(79) \\
6        & 0.02234(21) & 0.04336(71) \\
$\infty$ & 0.02088(39) & 0.0443(11)  \\
\hline
\end{tabular}
\end{center}
\caption{Continuum determinations of $\chi(N)$ obtained from the global continuum-large-$N$ combined extrapolation reported in Fig.~\ref{fig:largeN_GLOBAL}. These are my final results for the topological susceptibility. I've also expressed them in units of $\sqrt{8t_0}$ using the conversion factors in Tab.~\ref{tab:scale_conv}.}
\label{tab:chi_contlim_vs_N_global}
\end{table}

\noindent Compatible results within errors are obtained removing all $N=3$ data points and removing the $1/N^4$ term from the fit function. However, in order to be conservative, I will again quote the results of the quadratic fit in $N\ge3$, as they have larger statistical errors. Continuum determinations for $N=3,4,5,6,\infty$ obtained from the global fit are compared with those from the previously presented individual $N$-by-$N$ fits in Fig.~\ref{fig:comp_chi_MYRES}. As it can be appreciated, they perfectly agree within errors. In conclusion, I will quote the results from the global fit as my final results for the topological susceptibility for all $N$ values, similarly to Ref.~\cite{Ce:2016awn}. These are reported in Tab.~\ref{tab:chi_contlim_vs_N_global}. For the sake of completeness, I have also reported their values expressed in units of the popular scale $\sqrt{8t_0}$ using the conversion factors in Tab.~\ref{tab:scale_conv}.

\subsection{Discussion of the obtained results}\label{sec:res_discuss}

Before proceeding with the comparison of the results obtained thus far with those previously reported in the literature, I would like to draw a few interesting physical conclusions. First of all, I would like to discuss the conversion of my result to physical MeV units. Although such procedure is ambiguous in a non-physical theory such as large-$N$ Yang--Mills, given the phenomenological importance of $\chi_{\inf}$ it is worth to fix some prescription and still do it anyway. Mine will be to choose:
\beq
\sqrt{8t_0} \equiv 0.5~\mathrm{fm}, \qquad (N=\infty),
\eeq
since this is a popular choice in the large-$N$ literature adopting $t_0$ as scale setting. This choice leads to, cf.~Tab.~\ref{tab:scale_conv}:
\beq\label{eq:sigma_largeN}
\sqrt{\sigma} = 476~\mathrm{MeV}, \qquad (N=\infty),
\eeq
similar to and slightly larger than the recent lattice result found in $\Nf=2+1$ $N=3$ QCD for physical quark masses, $\sqrt{\sigma}=445(7)$ MeV~\cite{Bulava:2024jpj}.
This in turn yields:
\beq\label{eq:my_chi_inf_largeN_MeV}
\chi_{\inf} = \left[180.94(84)~\mathrm{MeV}\right]^{4}, \qquad (N=\infty).
\eeq
The value in~\eqref{eq:my_chi_inf_largeN_MeV} is perfectly compatible with the traditional result $\chi_{\inf}\simeq\left(180~\mathrm{MeV}\right)^4$ obtained from the evaluation of~\eqref{eq:witten_veneziano_formula} using the experimental values for $m_{\eta^\prime}$ and $F_\pi$. However, to avoid ambiguities due to the conversion to MeV units, let me also discuss the consistency of my large-$N$ result for the topological susceptibility with the Witten--Veneziano formula following a slightly different route, involving only dimensionless units.

Combining $\chi_{\inf}/\sigma^2 = 0.02088(39)$ with the recent large-$N$ continuum result $F_\pi/\sqrt{N\sigma} = 0.1262(34)$ of~\cite{Bonanno:2025hzr} (obtained by means of twisted volume reduction), I find:
\beq\label{eq:etaprime_largeN}
\sqrt{N} \, \frac{m_{\eta^\prime}}{\sqrt{\sigma}} \bigg\vert_{N\,=\,\infty} = \sqrt{2\Nf} \, \frac{\sqrt{\chi_{\inf}/\sigma^2}}{F_\pi/\sqrt{N\sigma}\big\vert_{N\,=\,\infty}} = 2.80(8),
\eeq
for the large-$N$ limit of $\sqrt{N}m_{\eta^\prime} \sim\mathcal{O}(N^0)$ in the $\Nf=3$ chiral limit. Using instead the experimental value $m_{\eta^\prime} = 957.78(6)$ MeV~\cite{ParticleDataGroup:2024cfk} and the lattice QCD result for physical quark masses $\sqrt{\sigma}=445(7)$ MeV~\cite{Bulava:2024jpj}, I obtain:
\beq\label{eq:etaprime_physicalpoint}
\sqrt{N} \, \frac{m_{\eta^\prime}}{\sqrt{\sigma}} \bigg\vert_{N\,=\,3} = 3.73(6).
\eeq
Clearly, these two numbers do not need to be equal, as they differ for finite-quark-mass and finite-$N$ corrections. However, they fall in the same ballpark, since
\beq\label{eq:subleading_corrs_etaprime}
\frac{\sqrt{N}m_{\eta^\prime}\big\vert_{N\,=\,3}}{\sqrt{N}m_{\eta^\prime}\big\vert_{N\,=\,\infty}} = 1.33(4),
\eeq
thus my result for $\chi_\infty$ is certainly large enough to support the Witten--Veneziano mechanism. From my previous experience with meson masses and decay constants in large-$N$ QCD~\cite{Bonanno:2023ypf,Bonanno:2025hzr}, I expect sub-leading corrections in $1/N$ between $N=\infty$ and $N=3$ to be much larger than finite-quark-mass ones between the physical point and the chiral limit, thus it is reasonable to expect that the deviation from 1 of the ratio in~\eqref{eq:subleading_corrs_etaprime} is mainly due to the former source. Given that $\chi_{\inf}$ in pure Yang--Mills theories receives leading $\mathcal{O}(1/N^2)$ corrections, one can naively expect that the size of $\mathcal{O}(1/N)$ sub-leading corrections to the estimate in Eq.~\eqref{eq:etaprime_largeN} should be of the same order of those affecting $F_\pi/\sqrt{N}$. These were recently computed in the continuum limit in Ref.~\cite{Bonanno:2025hzr}:
\beq
\frac{(F_{\pi}/\sqrt{N})\big\vert_{N\,=\,\infty}}{(F_{\pi}/\sqrt{N})\big\vert_{N,\,\Nf}} = 1 + 0.46(7)\frac{\Nf}{N} + \mathcal{O}\left(\frac{1}{N^2}\right).
\eeq
Interestingly, for $N=\Nf=3$, they agree in size and sign with~\eqref{eq:subleading_corrs_etaprime}, as I find $1.46(7)$. Although this is a crude estimate, this naive argument is consistent with a scenario where the difference between the large-$N$ and the physical point estimates~\eqref{eq:etaprime_largeN} and~\eqref{eq:etaprime_physicalpoint} is indeed due to sub-leading terms in the $1/N$ expansion.

\begin{figure}[!t]
\centering
\includegraphics[scale=0.5]{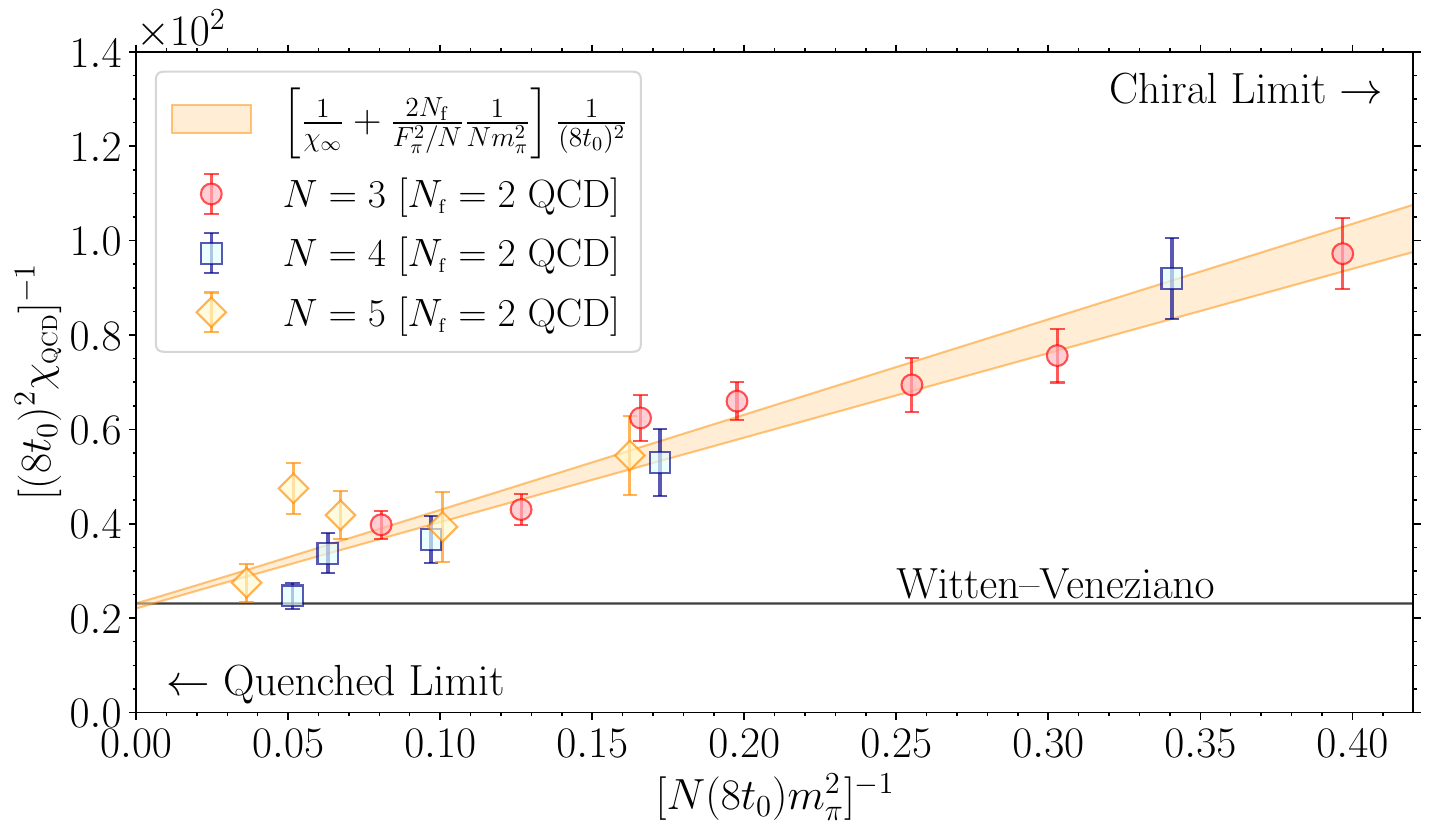}
\caption{Comparison of the large-$N$ lattice QCD data for the topological susceptibility obtained in~\cite{DeGrand:2020utq} with $\Nf=2$ flavors of dynamical Wilson quarks with the large-$N$ prediction in Eq.~\eqref{eq:chi_QCD_largeN}, obtained using my result for $\chi_{\inf}$ and the continuum large-$N$ result for $F_\pi/\sqrt{N}$ of~\cite{Bonanno:2025hzr}.}
\label{fig:chi_QCD_vs_YM}
\end{figure}

Another interesting large-$N$ topic I would like to discuss involving the Yang--Mills topological susceptibility is the following formula~\cite{Crewther:1977ce,DiVecchia:1980yfw,Leutwyler:1992yt}:
\beq\label{eq:chi_QCD_largeN}
\frac{1}{\chi_{\QCD}} = \frac{1}{\chi_{\inf}} + \frac{2\Nf}{F_\pi^2 m_\pi^2}.
\eeq
This equation expresses, at leading order in $1/N$, the non-commutativity of the large-$N$ and chiral limits for the topological susceptibility of full QCD $\chi_{\QCD}$, and it has the very interesting property of bridging the light-quark-mass and the heavy-quark-mass regimes for this observable. Indeed, when $m_\pi \to 0$ at fixed $N$ and $\Nf$, this equation reduces to the well-known formula for the topological susceptibility in Chiral Perturbation Theory: $\chi_\QCD=\frac{1}{2\Nf}F_\pi^2m_\pi^2$ (vanishing in the chiral limit). In the opposite quenched limit $m_\pi\to\infty$ (at fixed $N$ and $\Nf$) $\chi_{\QCD}$ tends instead to the pure-gauge result $\chi_{\inf}$, as it should since quarks decouple. Moreover, one similarly obtains $\chi_{\QCD} \to \chi_{\inf}$ taking $\Nf\to 0$, or taking $N\to\infty$ at fixed $m_\pi$ and $\Nf$, as $F_\pi^2 \sim \mathcal{O}(N)$.

Recently, a first lattice investigation with dynamical quarks of $\chi_\QCD$ in $\Nf=2$ QCD for $N=3,4,5$ in a wide range of pion masses appeared in~\cite{DeGrand:2020utq}. I will non-trivially check the consistency of Eq.~\eqref{eq:chi_QCD_largeN} by using these data to compute its left-hand side, and by obtaining its right-hand side from the combination of my result for $\chi_{\inf}/\sigma^2$ with the large-$N$ result of~\cite{Bonanno:2025hzr} for $F_\pi/\sqrt{N\sigma}$. Given that the author of~\cite{DeGrand:2020utq} uses $\sqrt{8t_0}$ to set the scale, I will suitably convert all large-$N$ quantities in these units using the $N=\infty$ result for $\sqrt{8t_0\sigma}$ in Tab.~\ref{tab:scale_conv}. The outcome of this comparison is shown in Fig.~\ref{fig:chi_QCD_vs_YM}. As it can by seen, there is a very good agreement between the full QCD lattice data for $1/\chi_{\QCD}$ as a function of $1/(N m_\pi^2)$ of~\cite{DeGrand:2020utq}, and the curve obtained computing $\frac{1}{\chi_{\inf}} + \frac{2N_{\scriptscriptstyle{\rm f}}}{F_\pi^2} \frac{1}{m_\pi^2}$ using my result for $\chi_{\inf}$ and the large-$N$ result for $F_\pi/\sqrt{N}$ of~\cite{Bonanno:2025hzr}. The only few data that deviate from the predicted curve are those obtained for the two largest $N$ at the heaviest pion masses. Given that these two regimes are known to suffer for much longer autocorrelation times when looking at the topological charge~\cite{Alles:1996vn,DelDebbio:2002xa,DelDebbio:2004xh,Schaefer:2010hu}, these deviations could be related to underestimated errors and/or incorrect sampling of $Q$ due to topological freezing.

\subsection{Comparison with previous determinations of $\chi$ in the literature}\label{sec:comp_chi}

\begin{figure}[!t]
\centering
\includegraphics[scale=0.5]{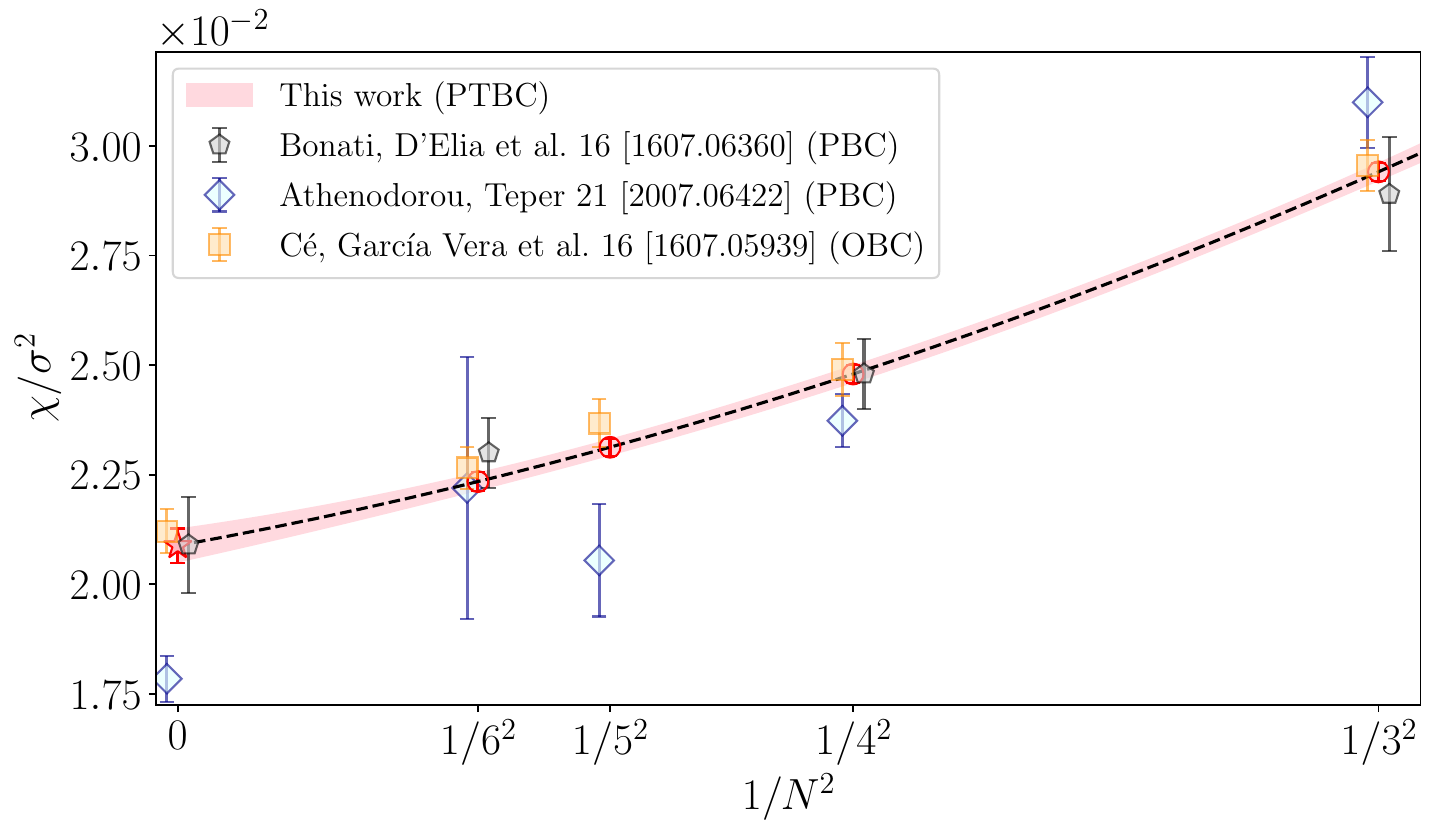}
\caption{Comparison of large-$N$ results for $\chi(N)$ found in the literature, see Sec.~\ref{sec:comp_chi}.}
\label{fig:chi_vs_N_ALL_COMP}
\end{figure}

\begin{figure}[!t]
\centering
\includegraphics[scale=0.6]{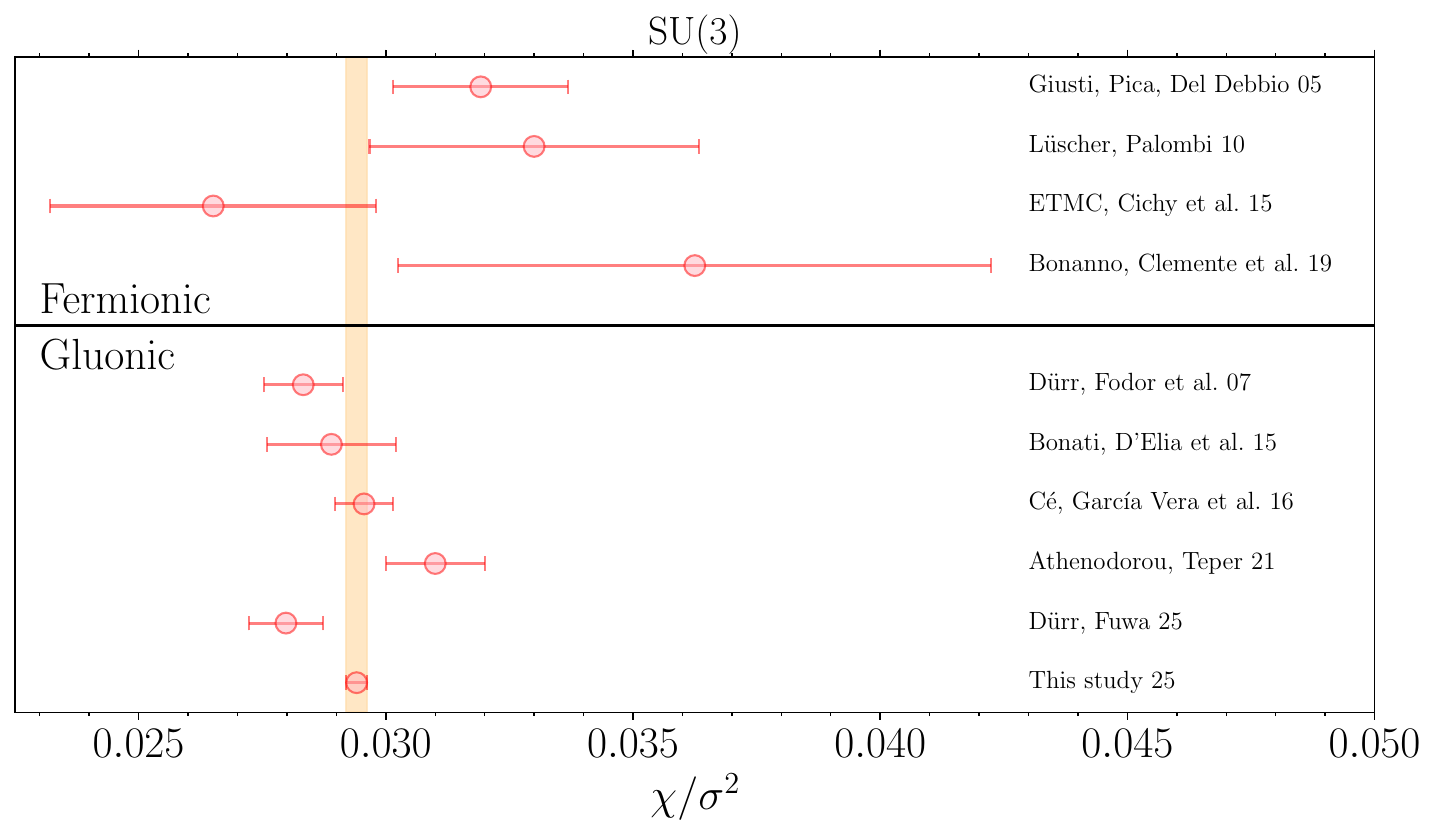}
\caption{Summary plot of the determinations of the topological susceptibility $\chi/\sigma^2$ for $N=3$, see also the related Tab.~\ref{tab:summary_N3}.}
\label{fig:summary_N3}
\end{figure}

\begin{figure}[!t]
\centering
\includegraphics[scale=0.6]{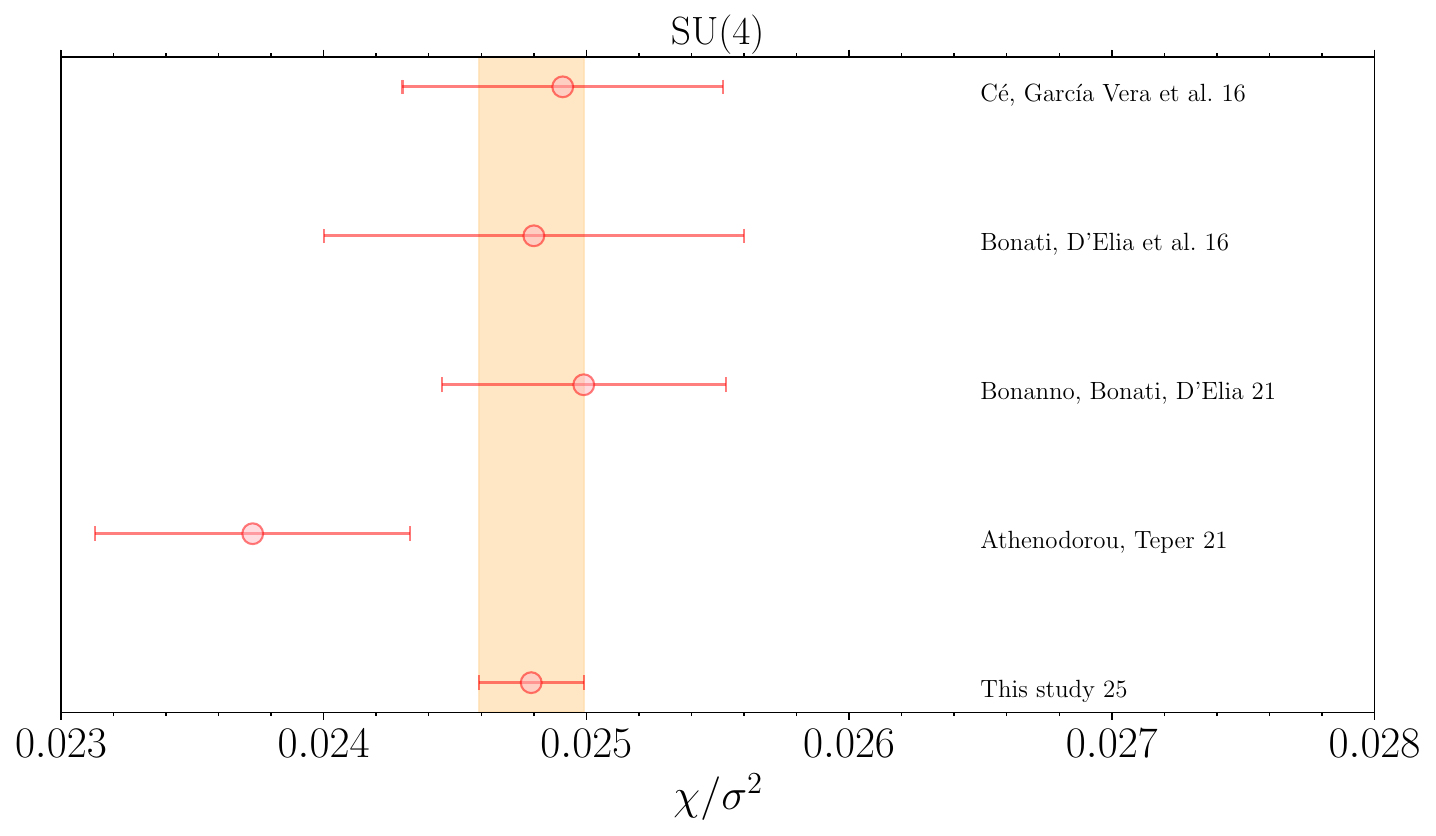}
\caption{Summary plot of the determinations of the topological susceptibility $\chi/\sigma^2$ for $N=4$, see also the related Tab.~\ref{tab:summary_N4}.}
\label{fig:summary_N4}
\end{figure}

\begin{figure}[!t]
\centering
\includegraphics[scale=0.6]{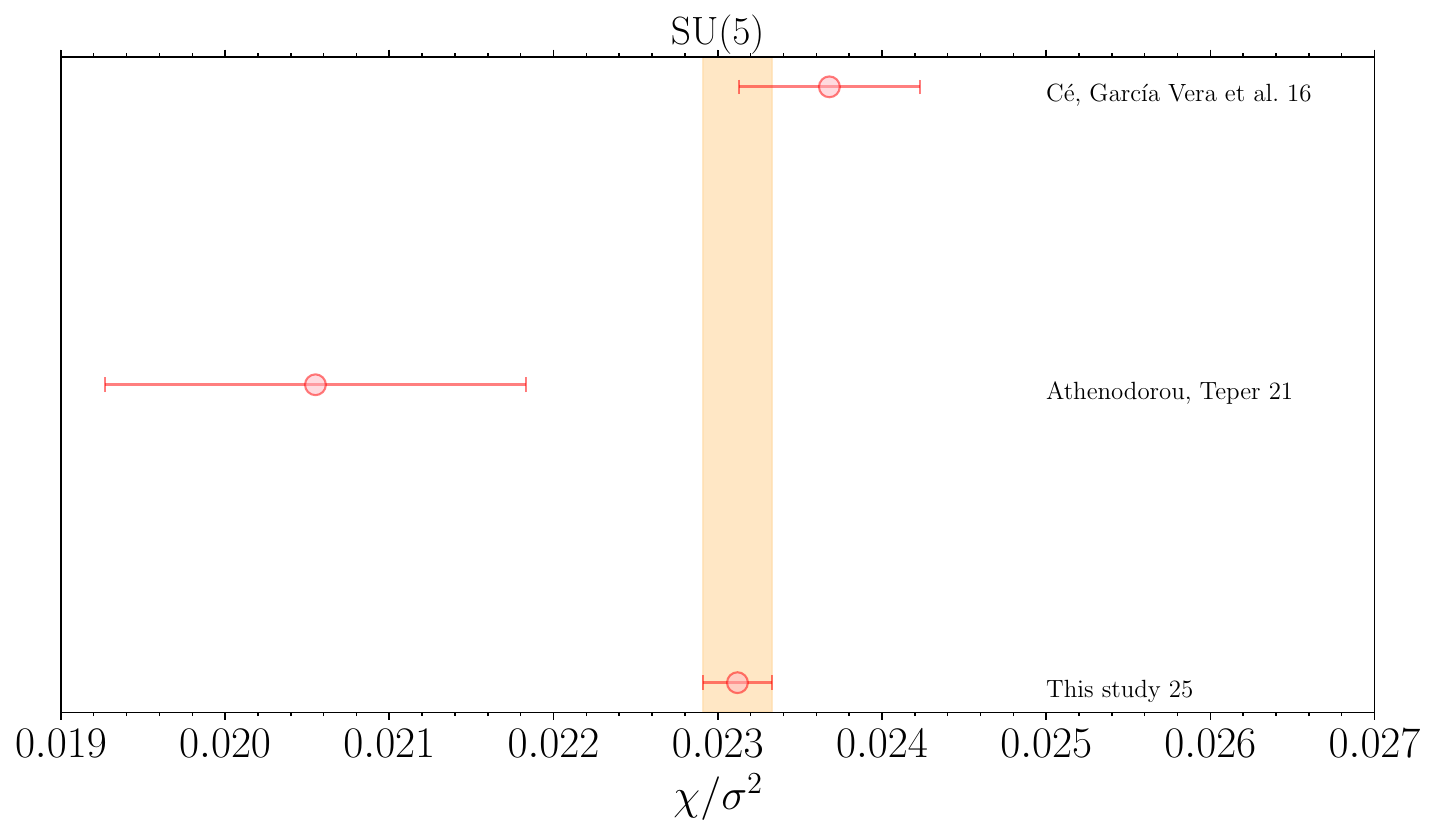}
\caption{Summary plot of the determinations of the topological susceptibility $\chi/\sigma^2$ for $N=5$, see also the related Tab.~\ref{tab:summary_N5}.}
\label{fig:summary_N5}
\end{figure}

\begin{figure}[!t]
\centering
\includegraphics[scale=0.6]{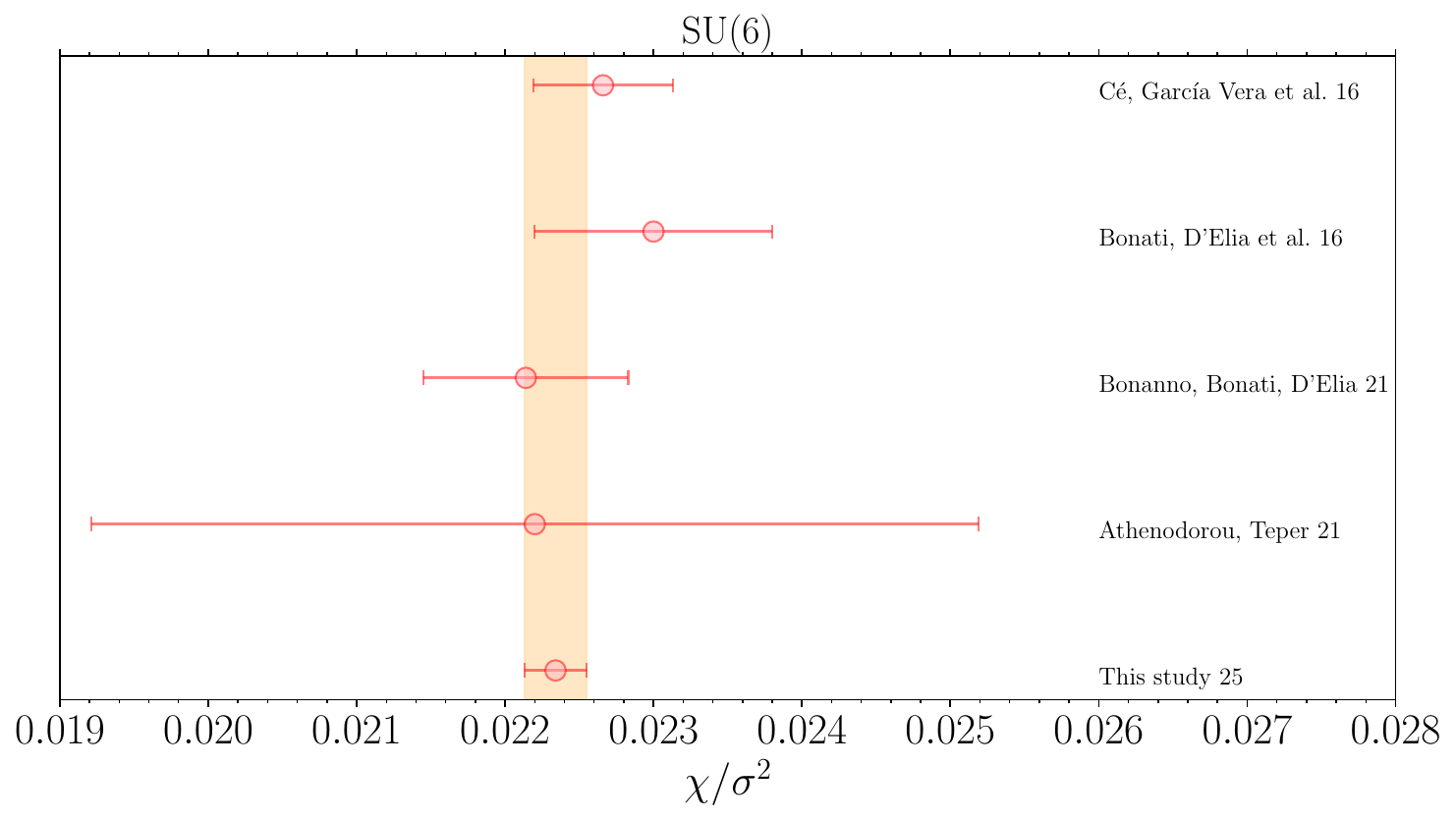}
\caption{Summary plot of the determinations of the topological susceptibility $\chi/\sigma^2$ for $N=6$, see also the related Tab.~\ref{tab:summary_N6}.}
\label{fig:summary_N6}
\end{figure}

\begin{figure}[!t]
\centering
\includegraphics[scale=0.6]{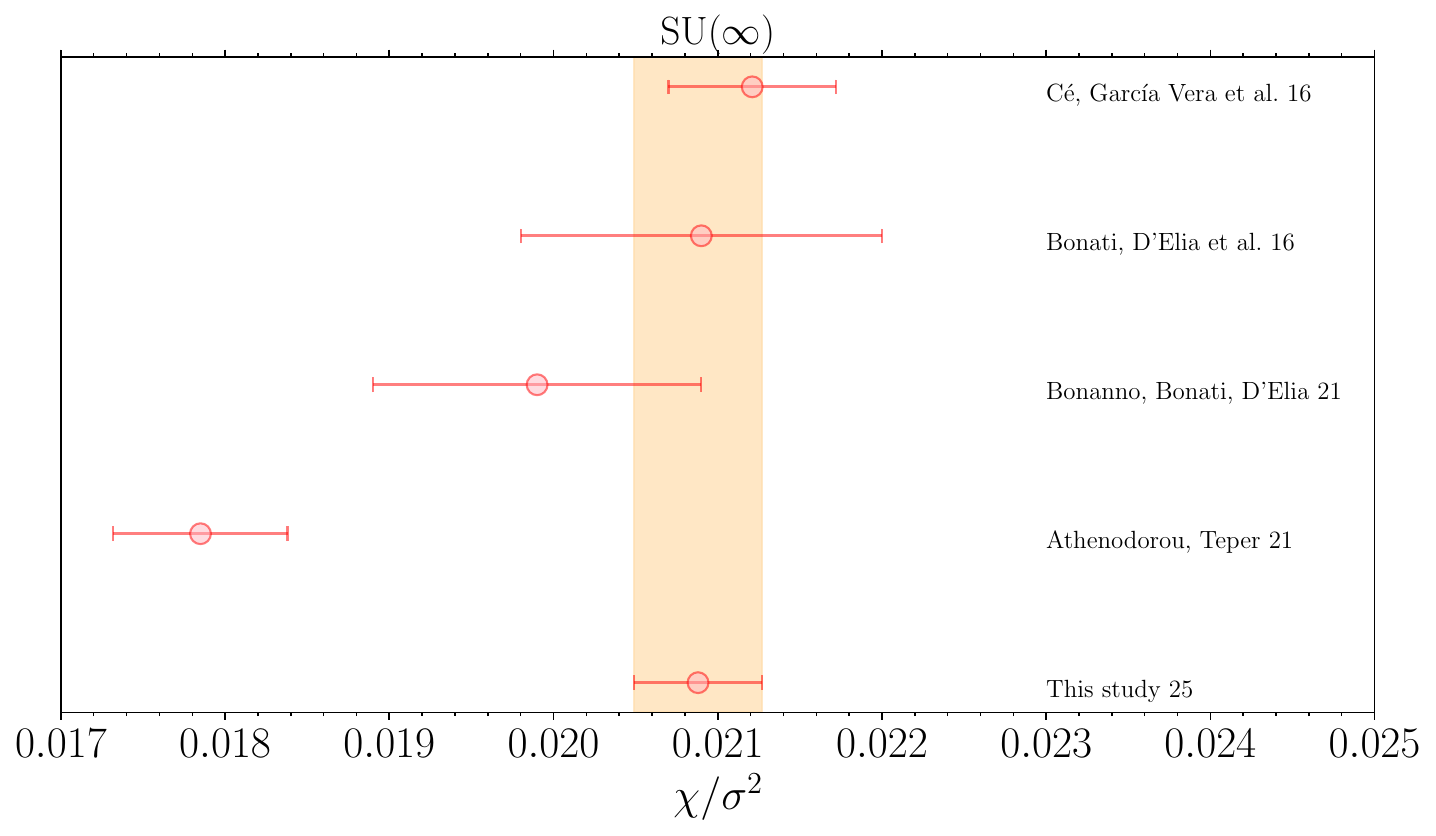}
\caption{Summary plot of the determinations of the topological susceptibility $\chi/\sigma^2$ for $N=\infty$, see also the related Tab.~\ref{tab:summary_N_INF}.}
\label{fig:summary_N_INF}
\end{figure}

I will now conclude my investigation with a comprehensive comparison of several determinations of $\chi$ in pure $\SU(N)$ Yang--Mills theories with my new results. All previous findings expressed in units of $r_0$ or $\sqrt{8t_0}$ have been converted in units of the string tension $\sqrt{\sigma}$ via the continuum conversion factors in Tab.~\ref{tab:scale_conv}. The criterion I have followed to gather data is to consider only those references reporting continuum-extrapolated results.\\
I show all determinations of $\chi$ as a function of $N$ retrieved in the literature in the collective Fig.~\ref{fig:chi_vs_N_ALL_COMP}, which only quotes works presenting results for $N\ge 3$ with different algorithms than PTBC. A more complete collection of results for $\chi$ is gathered in Tabs.~\ref{tab:summary_N3}, \ref{tab:summary_N4}, \ref{tab:summary_N5}, \ref{tab:summary_N6}, \ref{tab:summary_N_INF} for $N=3,4,5,6,\infty$ respectively. Such data are also shown in Figs.~\ref{fig:summary_N3}, \ref{fig:summary_N4}, \ref{fig:summary_N5}, \ref{fig:summary_N6}, \ref{fig:summary_N_INF}. For illustrative purposes, tables include naive weighted averages of collected data with their relative chi-squared and $p$-values. In Figs.~\ref{fig:summary_N3}--\ref{fig:summary_N_INF}, shaded bands correspond to my new PTBC determinations, and illustrate how they compare with previous ones.

\newpage

\FloatBarrier

\begin{table}[!t]
\begin{center}
\begin{tabular}{|c|c|c|}
\hline
\multicolumn{3}{|c|}{$N=3$} \\
\hline
\hline
$\chi/\sigma^2$ & Tag and Reference & Details \\
\hline
\hline
\multicolumn{3}{|c|}{\textbf{Gluonic}} \\
\hline
0.02941(21)  & This study 25                                     & Cooling \\
0.02798(75)  & D{\"u}rr, Fuwa 25~\cite{Durr:2025qtq}             & Gradient flow/Stout smearing  \\
0.03100(100) & Athenodorou, Teper 20~\cite{Athenodorou:2020ani}  & Cooling \\
0.02956(58)  & Cé, Garc\'ia Vera et al.~15~\cite{Ce:2015qha}     & Gradient flow \\
0.02890(130) & Bonati, D'Elia et al.~15~\cite{Bonati:2015sqt}    & Cooling        \\
0.02833(80)  & D{\"u}rr, Fodor et al.~07~\cite{Durr:2006ky}     & HYP Smearing \\
\hline
\hline
\multicolumn{3}{|c|}{\textbf{Fermionic}}\\
\hline
0.03625(600) & Bonanno, Clemente et al.~19~\cite{Bonanno:2019xhg}  & Staggered \\
0.02651(330) & ETMC, Cichy et al.~15~\cite{Cichy:2015jra}          & \makecell{Wilson (Iwasaki gauge action)}\\
0.03300(333) & L{\"u}scher, Palombi 10~\cite{Luscher:2010ik}       & Wilson \\
0.03192(177) & Giusti, Pica, Del Debbio 05~\cite{DelDebbio:2004ns} & Overlap \\
\hline
\hline
\multicolumn{3}{|c|}{\textbf{Average}: 0.02936(18) $\,\qquad$[reduced chi-squared: 1.49, $p$-value $=14.6\%$]} \\
\hline
\end{tabular}
\end{center}
\caption{Summary of continuum determinations of the topological susceptibility $\chi/\sigma^2$ for the $\SU(3)$ pure Yang--Mills theory. The central column translates tags appearing in Fig.~\ref{fig:summary_N3} to the corresponding references in the bibliography. All studies adopted the standard Wilson gauge action unless stated otherwise, and all assumed periodic boundaries.}
\label{tab:summary_N3}
\end{table}

\FloatBarrier

\begin{table}[!t]
\begin{center}
\begin{tabular}{|c|c|c|}
\hline
\multicolumn{3}{|c|}{$N=4$} \\
\hline
\hline
$\chi/\sigma^2$ & Tag and Reference & Details \\
\hline
\hline
0.02479(20)  & This study 25                                     & Cooling, PTBC \\
0.02373(60)  & Athenodorou, Teper 21~\cite{Athenodorou:2021qvs}  & Cooling, PBC \\
0.02499(54)  & Bonanno, Bonati, D'Elia~21~\cite{Bonanno:2020hht} & Cooling, PTBC \\
0.02480(80)  & Bonati, D'Elia et al.~16~\cite{Bonati:2016tvi}    & Cooling, PBC  \\
0.02491(61)  & Cé, Garc\'ia Vera et al.~16~\cite{Ce:2016awn}     & Gradient flow, OBC \\
\hline
\hline
\multicolumn{3}{|c|}{\textbf{Average}: 0.02473(17) $\,\qquad$[reduced chi-squared: 0.80, $p$-value $=52.6\%$]} \\
\hline
\end{tabular}
\end{center}
\caption{Summary of continuum determinations of the topological susceptibility $\chi/\sigma^2$ for the $\SU(4)$ pure Yang--Mills theory. The central column translates tags appearing in Fig.~\ref{fig:summary_N4} to the corresponding references in the bibliography. All studies adopted the standard Wilson gauge action.}
\label{tab:summary_N4}
\end{table}

\FloatBarrier

\begin{table}[!t]
\begin{center}
\begin{tabular}{|c|c|c|}
\hline
\multicolumn{3}{|c|}{$N=5$} \\
\hline
\hline
$\chi/\sigma^2$ & Tag and Reference & Details \\
\hline
\hline
0.02312(21)   & This study 25                                     & Cooling, PTBC \\
0.02055(128)* & Athenodorou, Teper 21~\cite{Athenodorou:2021qvs}  & Cooling, PBC \\
0.02368(55)   & Cé, Garc\'ia Vera et al.~16~\cite{Ce:2016awn}     & Gradient flow, OBC \\
\hline
\hline
\multicolumn{3}{|c|}{\textbf{Average} (no *): 0.02319(20) $\,\qquad$[reduced chi-squared: 0.90, $p$-value $=34.2\%$]} \\
\hline
\hline
\multicolumn{3}{|c|}{\textbf{Average} (with *): 0.02313(19) $\,\qquad$[reduced chi-squared: 2.5, $p$-value $=7.9\%$]} \\
\hline
\end{tabular}
\end{center}
\caption{Summary of continuum determinations of the topological susceptibility $\chi/\sigma^2$ for the $\SU(5)$ pure Yang--Mills theory. The central column translates tags appearing in Fig.~\ref{fig:summary_N5} to the corresponding references in the bibliography. All studies adopted the standard Wilson gauge action.}
\label{tab:summary_N5}
\end{table}

\FloatBarrier

\begin{table}[!t]
\begin{center}
\begin{tabular}{|c|c|c|}
\hline
\multicolumn{3}{|c|}{$N=6$} \\
\hline
\hline
$\chi/\sigma^2$ & Tag and Reference & Details \\
\hline
\hline
0.02234(21)  & This study 25                                     & Cooling, PTBC \\
0.02220(299) & Athenodorou, Teper 21~\cite{Athenodorou:2021qvs}  & Cooling, PBC \\
0.02214(69)  & Bonanno, Bonati, D'Elia~21~\cite{Bonanno:2020hht} & Cooling, PTBC \\
0.02300(80)  & Bonati, D'Elia et al.~16~\cite{Bonati:2016tvi}    & Cooling, PBC  \\
0.02266(47)  & Cé, Garc\'ia Vera et al.~16~\cite{Ce:2016awn}     & Gradient flow, OBC \\
\hline
\hline
\multicolumn{3}{|c|}{\textbf{Average}: 0.02241(18) $\,\qquad$[reduced chi-squared: 0.27, $p$-value $=89.5\%$]} \\
\hline
\end{tabular}
\end{center}
\caption{Summary of continuum determinations of the topological susceptibility $\chi/\sigma^2$ for the $\SU(6)$ pure Yang--Mills theory. The central column translates tags appearing in Fig.~\ref{fig:summary_N6} to the corresponding references in the bibliography. All studies adopted the standard Wilson gauge action.}
\label{tab:summary_N6}
\end{table}

\FloatBarrier

\begin{table}[!t]
\begin{center}
\begin{tabular}{|c|c|c|}
\hline
\multicolumn{3}{|c|}{$N=\infty$} \\
\hline
\hline
$\chi/\sigma^2$ & Tag and Reference & Details \\
\hline
\hline
0.02088(39)  & This study 25                                     & Cooling, PTBC \\
0.01785(53)* & Athenodorou, Teper 21~\cite{Athenodorou:2021qvs}  & Cooling, PBC \\
0.01990(100) & Bonanno, Bonati, D'Elia~21~\cite{Bonanno:2020hht} & Cooling, PTBC \\
0.02090(110) & Bonati, D'Elia et al.~16~\cite{Bonati:2016tvi}    & Cooling, PBC  \\
0.02121(51)  & Cé, Garc\'ia Vera et al.~16~\cite{Ce:2016awn}     & Gradient flow, OBC \\
\hline
\hline
\multicolumn{3}{|c|}{\textbf{Average} (no *): 0.02090(29) $\,\qquad$[reduced chi-squared: 0.46, $p$-value $=71.2\%$]} \\
\hline
\hline
\multicolumn{3}{|c|}{\textbf{Average} (with *): 0.02020(25) $\,\qquad$[reduced chi-squared: 6.78, $p$-value $=0.002\%$]} \\
\hline
\end{tabular}
\end{center}
\caption{Summary of continuum determinations of the topological susceptibility $\chi/\sigma^2$ for the $\SU(\infty)$ pure Yang--Mills theory. The central column translates tags appearing in Fig.~\ref{fig:summary_N_INF} to the corresponding references in the bibliography. All studies adopted the standard Wilson gauge action.}
\label{tab:summary_N_INF}
\end{table}

\FloatBarrier

The case $N=3$ has the larger number of entries, and is the only case where I could retrieve determinations obtained both via gluonic definitions~\cite{Durr:2006ky,Bonati:2015sqt,Ce:2015qha,Athenodorou:2020ani,Durr:2025qtq}---computed employing several smoothing algorithms---and fermionic ones~\cite{DelDebbio:2004ns,Luscher:2010ik,Cichy:2015jra,Bonanno:2019xhg}---with calculations performed adopting both chiral and non-chiral quark discretizations. Due to the much larger computational burden, fermionic definitions are systematically less precise compared to gluonic ones. Modulo this caveat, there is overall a good agreement between fermionic and gluonic results, and among all results as a whole. A naive constant fit to the full data set yields a reduced chi-squared of about 1.5 with 9 degrees of freedom, corresponding to a $\sim 15\%$ $p$-value, indeed confirming the observed agreement.

\begin{figure}[!t]
\centering
\includegraphics[scale=0.48]{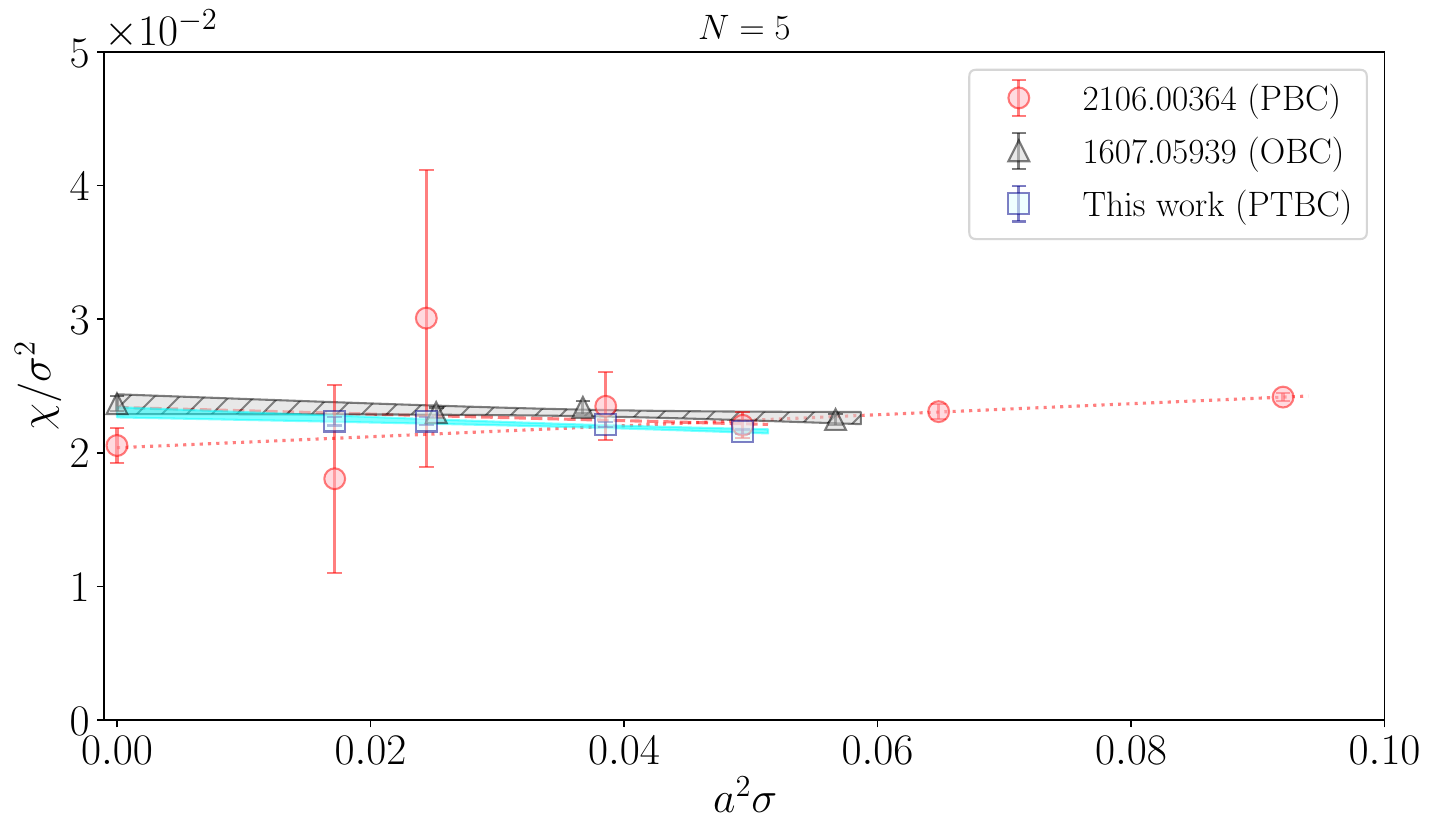}
\caption{Comparison of the data for $\chi/\sigma^2$ for the SU(5) Yang--Mills theory obtained in this paper and in Ref.~\cite{Athenodorou:2021qvs}. I also show the continuum limit of Ref.~\cite{Ce:2016awn}.}
\label{fig:teper_check_N5}
\end{figure}

The case $N>3$ has been less investigated, and only gluonic calculations have been performed. In these cases, on one hand I observe very good agreement among results obtained with the PTBC algorithm, Ref.~\cite{Bonanno:2020hht} and the current results, with OBC~\cite{Ce:2016awn}, and the first continuum determinations obtained with PBC~\cite{Bonati:2016tvi}. On the other hand, the results of~\cite{Athenodorou:2021qvs} for $N=4$ and $N=5$, obtained with PBC as well, are in tension with all the other ones, and systematically smaller.\footnote{In~\cite{Athenodorou:2021qvs}, the authors report two different determinations of the topological susceptibility: one is obtained with the same definition of the lattice clover cooled charge adopted here, the other one rounding $Q_{\L}$ to the nearest integer. Given that the authors found perfect agreement among the two determinations in the continuum limit (see also~\cite{Bonati:2015sqt} in this respect), for the purpose of this comparison, I have only considered the former, being it the same I have also employed in my calculations.} The largest tension is seen for $N=5$ (as an example, performing a constant fit to all data, this would yield a large reduced chi-squared $\sim 2.5$ and a poor $p$-value of $\sim 7\%$). The origin of this tension seems to be due to the inclusion of too coarse lattice spacings $a^2\sigma > 0.06$ in Ref.~\cite{Athenodorou:2021qvs}'s continuum limit, as they drive the extrapolation towards much smaller values. Restricting the extrapolation range to $a^2\sigma < 0.06$, I instead observe that the continuum extrapolation of Ref.~\cite{Athenodorou:2021qvs}'s data for $N=5$ would be perfectly compatible with mine and with Ref.~\cite{Ce:2016awn}'s one, see Fig.~\ref{fig:teper_check_N5}.

Finally, concerning $N=\infty$ extrapolations, the situation is similar to the $N=4,5$ cases. Indeed, my new PTBC result agrees very well with previous large-$N$ extrapolations obtained with PBC~\cite{Bonati:2016tvi} and OBC~\cite{Ce:2016awn}, as well as with the previous PTBC result of~\cite{Bonanno:2020hht}, while the result of~\cite{Athenodorou:2021qvs} lies much below. This is not surprising, given that already Ref.~\cite{Athenodorou:2021qvs}'s results for $N=4$ and $N=5$ are systematically smaller compared to the others, and completely dominate the large-$N$ extrapolation due to the very larger error on~\cite{Athenodorou:2021qvs}'s result for $N=6$. Moreover, it should be noted that the authors of~\cite{Athenodorou:2021qvs} also included $N=2$ and $3$ data in their large-$N$ extrapolation without including any higher-order term in $1/N^2$. This is risky, as my results strongly point out that these lie outside of the region where sub-leading corrections to $N=\infty$ can be described by a simple linear term in $1/N^2$. Thus, this element too could bias the extrapolation towards smaller values.

In conclusions, my new PTBC determinations are generally in very good agreement, both for finite number of colors and in the large-$N$ limit, with previous determinations in the literature. The observed agreement is thus a very rewarding achievement, given the very different numerical strategies, algorithms and lattice discretizations involved. The only exceptions are the few cases earlier discussed concerning Ref.~\cite{Athenodorou:2021qvs}, where however the origin of the discrepancy has been traced back, and is likely related to extrapolations rather than to the lattice data themselves.

\section{Conclusions}\label{sec:conclu}

In this paper I have presented a dedicated study of the large-$N$ limit of the topological susceptibility of $\SU(N)$ Yang--Mills theories, performed adopting the Parallel Tempering on Boundary Conditions (PTBC) algorithm to avoid topological freezing at large $N$.

I have presented results for the autocorrelation times of PTBC and discussed their scaling with the lattice spacing and with the number of colors, showing the improvement achieved with respect to standard algorithms. The adoption of PTBC permitted me to determine the topological susceptibility with high accuracy for finer lattice spacings compared to previous investigations with periodic and open boundary conditions, allowing me to achieve controlled continuum limits of $\chi$ for all $N$. By taking the continuum limit of $\chi$ at fixed smoothing radius in physical units for several choices of $\Rs$, I also showed the independence of $\chi$ on $\Rs$ in the continuum limit in the probed range of smoothing radii. Combining my extrapolated result for the large-$N$ limit of $\chi$ with a recent large-$N$ determination of $F_\pi$, I have found excellent agreement both with the Witten--Veneziano formula, and with the Di Vecchia--Veneziano equation describing the relation between quenched and unquenched $\chi$ at large $N$. Finally, I have performed a comprehensive comparison of my determinations with previous ones in the literature, generally finding very good agreement with previous PBC and OBC calculations, and clarifying the origin of a few tensions with Ref.~\cite{Athenodorou:2021qvs}'s results. The most important results presented in this manuscript are summarized in the text box below:

\begin{center}
\begin{tcolorbox}[enhanced,width=\textwidth,center upper,size=fbox,
fontupper=\large\bfseries,sharp corners]
\beq
\frac{1}{\sigma^2}\chi(N) = 0.02088(39) + 0.044(12) \frac{1}{N^2} + 0.293(83)\frac{1}{N^4} + \mathcal{O}\left(\frac{1}{N^6}\right),
\eeq
\beq
\sqrt{N} \, \frac{m_{\eta^\prime}}{\sqrt{\sigma}}\bigg\vert_{N\,=\,\infty} = 2.80(8),
\eeq
\beq
\frac{\sigma^2}{\chi_{\QCD}} = 47.9(9) + 126(7) \frac{\Nf}{N} \frac{\sigma}{m_\pi^2}.
\eeq
\end{tcolorbox}
\end{center}

Overall, the results presented in this paper show once again the reliability of the PTBC algorithm in improving the efficiency of lattice simulations of Yang--Mills theories for what concerns the mitigation of topological freezing. Given that this algorithm was recently implemented with dynamical fermions~\cite{Bonanno:2024zyn}, it is natural to look forward to new applications of the PTBC in several QCD physical contexts where topological freezing hinders progress, such as the study of the QCD topological susceptibility~\cite{Borsanyi:2016ksw,Petreczky:2016vrs,Bonati:2018blm,Athenodorou:2022aay,Kotov:2025ilm} or of the QCD sphaleron rate at very high temperatures~\cite{Bonanno:2023ljc, Bonanno:2023thi}, which are very prominent topics in relation with Standard Model~\cite{Fukushima:2008xe} and Beyond Standard Model~\cite{Notari:2022ffe} phenomenology. Also the study of $\chi$ at large $N$ with dynamical fermions~\cite{DeGrand:2020utq} could greatly benefit from the adoption of PTBC and would be an intriguing future outlook. Finally, it would be very interesting to use PTBC to push forward the study of the topological susceptibility slope $\chi^\prime$ I have performed in SU(3)~\cite{Bonanno:2023ple} by extending it to the large-$N$ case, as this topic has both theoretical and phenomenological importance~\cite{Tarasov:2025mvn}.

\section*{Acknowledgements}
I am grateful to C.~Bonati, M.~D'Elia, and M. Garc\'ia P\'erez for useful discussions and comments, and for reading this manuscript. This work is supported by the Spanish Research Agency (Agencia Estatal de Investigaci\'on) through the grant IFT Centro de Excelencia Severo Ochoa CEX2020-001007-S and, partially, by the grant PID2021-127526NB-I00, both funded by MCIN/AEI/10.13039/501100011033. Numerical calculations have been performed on the \texttt{Finisterrae~III} cluster at CESGA (Centro de Supercomputaci\'on de Galicia).

\FloatBarrier
\newpage
\appendix
\section*{Appendix}
\section{Raw data}\label{app:rawdata}

In this appendix I have reported all raw data collected for the topological susceptibility for several values of $N$, $\beta$ and $n_\cool$: Tabs.~\ref{tab:rawdata_N3},~\ref{tab:rawdata_N4},~\ref{tab:rawdata_N5},~\ref{tab:rawdata_N6} collect all data for, respectively, $N=3,4,5,6$.

\FloatBarrier

\begin{table}[!htb]
\tiny
\begin{center}
\begin{tabular}{|cc|cc|cc|cc|cc|}
\hline
\multicolumn{10}{|c|}{$N=3$}\\
\hline
\hline
\multicolumn{10}{|c|}{$10^6 \times a^4 \chi_{\L}(n_\cool)$}\\
\hline
\hline
\multicolumn{2}{|c}{$\beta=5.95$} &\multicolumn{2}{|c}{$\beta=6.00$}&\multicolumn{2}{|c}{$\beta=6.07$}&\multicolumn{2}{|c}{$\beta=6.20$}&\multicolumn{2}{|c|}{$\beta=6.40$}\\
\hline
\hline
2  &  60.17(8)   &  2  &  43.81(7)  &  2  &  28.41(6)  &  3  &  14.76(8)  &  5  &  5.16(3) \\
3  &  69.76(10)  &  4  &  53.72(9)  &  4  &  34.30(8)  &  6  &  16.42(9)  &  10 &  5.47(3) \\
4  &  74.79(11)  &  6  &  57.33(10) &  6  &  36.44(8)  &  9  &  17.04(9)  &  15 &  5.59(3) \\
5  &  77.96(11)  &  8  &  59.27(10) &  8  &  37.57(9)  &  12 &  17.37(9)  &  20 &  5.65(3) \\
6  &  80.16(11)  &  10 &  60.48(10) &  10 &  38.29(9)  &  15 &  17.58(9)  &  25 &  5.69(3) \\
7  &  81.78(12)  &  12 &  61.31(10) &  12 &  38.78(9)  &  18 &  17.73(9)  &  30 &  5.72(3) \\
8  &  83.01(12)  &  14 &  61.92(10) &  14 &  39.14(9)  &  21 &  17.84(9)  &  35 &  5.74(3) \\
9  &  83.99(12)  &  16 &  62.39(10) &  16 &  39.43(9)  &  24 &  17.92(9)  &  40 &  5.75(3) \\
10 &  84.79(12)  &  18 &  62.76(11) &  18 &  39.65(9)  &  27 &  17.99(9)  &  45 &  5.77(3) \\
11 &  85.44(12)  &  20 &  63.06(11) &  20 &  39.83(9)  &  30 &  18.05(9)  &  50 &  5.78(3) \\
12 &  85.99(12)  &  22 &  63.30(11) &  22 &  39.99(9)  &  33 &  18.09(10) &  55 &  5.79(3) \\
13 &  86.46(12)  &     &            &  24 &  40.12(9)  &  36 &  18.13(10) &  60 &  5.80(3) \\
14 &  86.87(12)  &     &            &  26 &  40.23(9)  &  39 &  18.17(10) &  65 &  5.80(3) \\
15 &  87.22(12)  &     &            &  28 &  40.32(9)  &  42 &  18.20(10) &  70 &  5.81(3) \\
16 &  87.53(12)  &     &            &     &            &     &            &  75 &  5.81(3) \\
17 &  87.81(12)  &     &            &     &            &     &            &     &          \\
18 &  88.05(12)  &     &            &     &            &     &            &     &          \\
19 &  88.27(13)  &     &            &     &            &     &            &     &          \\
\hline
\end{tabular}
\end{center}
\caption{Each column reports $10^6 \times a^4\chi_{\L}$ as a function of $n_\cool$ for a given $\beta$. Data refer to $N=3$.}
\label{tab:rawdata_N3}
\end{table}

\begin{table}[!htb]
\tiny
\begin{center}
\begin{tabular}{|cc|cc|cc|cc|}
\hline
\multicolumn{8}{|c|}{$N=4$}\\
\hline
\hline
\multicolumn{8}{|c|}{$10^6 \times a^4 \chi_{\L}(n_\cool)$}\\
\hline
\hline
\multicolumn{2}{|c}{$\beta=11.02$} &\multicolumn{2}{|c}{$\beta=11.20$}&\multicolumn{2}{|c}{$\beta=11.40$}&\multicolumn{2}{|c|}{$\beta=11.60$}\\
\hline
\hline
2  &  35.67(14)  &  2  &  18.90(9)   &  3  &  11.01(9)   &  4 &  6.12(6) \\
3  &  40.92(16)  &  4  &  22.63(11)  &  6  &  12.22(11)  &  8 &  6.59(6) \\
4  &  43.62(17)  &  6  &  23.97(12)  &  9  &  12.67(11)  &  12 &  6.76(7) \\
5  &  45.32(18)  &  8  &  24.70(12)  &  12 &  12.92(11)  &  16 &  6.86(7) \\
6  &  46.50(19)  &  10 &  25.16(12)  &  15 &  13.08(11)  &  20 &  6.92(7) \\
7  &  47.38(19)  &  12 &  25.48(12)  &  18 &  13.19(11)  &  24 &  6.96(7) \\
8  &  48.07(19)  &  14 &  25.73(12)  &  21 &  13.27(11)  &  28 &  7.00(7) \\
9  &  48.62(19)  &  16 &  25.92(12)  &  24 &  13.34(12)  &  32 &  7.02(7) \\
10 &  49.08(20)  &  18 &  26.07(13)  &  27 &  13.39(12)  &  36 &  7.04(7) \\
11 &  49.46(20)  &  20 &  26.20(13)  &  30 &  13.43(12)  &  40 &  7.06(7) \\
12 &  49.79(20)  &  22 &  26.31(13)  &  33 &  13.47(12)  &  44 &  7.07(7) \\
13 &  50.08(20)  &  24 &  26.41(13)  &  36 &  13.50(12)  &  48 &  7.09(7) \\
14 &  50.33(20)  &  26 &  26.49(13)  &  39 &  13.53(12)  &  52 &  7.10(7) \\
15 &  50.55(20)  &  28 &  26.56(13)  &  42 &  13.56(12)  &  56 &  7.11(7) \\
16 &  50.75(20)  &  30 &  26.63(13)  &  45 &  13.58(12)  &  60 &  7.11(7) \\
17 &  50.93(20)  &  32 &  26.68(13)  &     &             &  64 &  7.12(7) \\
18 &  51.10(20)  &     &             &     &             &     &          \\
19 &  51.25(20)  &     &             &     &             &     &          \\
20 &  51.39(20)  &     &             &     &             &     &          \\
21 &  51.51(21)  &     &             &     &             &     &          \\
\hline
\end{tabular}
\end{center}
\caption{Each column reports $10^6 \times a^4\chi_{\L}$ as a function of $n_\cool$ for a given $\beta$. Data refer to $N=4$.}
\label{tab:rawdata_N4}
\end{table}

\FloatBarrier

\begin{table}[!htb]
\tiny
\begin{center}
\begin{tabular}{|cc|cc|cc|cc|}
\hline
\multicolumn{8}{|c|}{$N=5$}\\
\hline
\hline
\multicolumn{8}{|c|}{$10^6 \times a^4 \chi_{\L}(n_\cool)$}\\
\hline
\hline
\multicolumn{2}{|c}{$\beta=17.43$} &\multicolumn{2}{|c}{$\beta=17.63$}&\multicolumn{2}{|c}{$\beta=18.04$}&\multicolumn{2}{|c|}{$\beta=18.375$}\\
\hline
\hline
2 &  37.71(11)  &  2 &  23.77(8)  &  3 &  11.00(8)  &  3 &  5.49(7) \\
3 &  43.35(13)  &  4 &  28.68(10)  &  6 &  12.23(9)  &  6 &  6.04(7) \\
4 &  46.23(14)  &  6 &  30.45(11)  &  9 &  12.68(10)  &  9 &  6.25(7) \\
5 &  48.04(15)  &  8 &  31.40(11)  &  12 &  12.93(10)  &  12 &  6.36(8) \\
6 &  49.30(15)  &  10 &  32.02(11)  &  15 &  13.09(10)  &  15 &  6.43(8) \\
7 &  50.25(15)  &  12 &  32.45(11)  &  18 &  13.20(10)  &  18 &  6.48(8) \\
8 &  50.98(16)  &  14 &  32.77(12)  &  21 &  13.29(10)  &  21 &  6.52(8) \\
9 &  51.57(16)  &  16 &  33.03(12)  &  24 &  13.35(10)  &  24 &  6.55(8) \\
10 &  52.06(16)  &  18 &  33.24(12)  &  27 &  13.40(10)  &  27 &  6.57(8) \\
11 &  52.47(16)  &  20 &  33.41(12)  &  30 &  13.45(10)  &  30 &  6.59(8) \\
12 &  52.83(16)  &  22 &  33.56(12)  &  33 &  13.49(10)  &  33 &  6.61(8) \\
13 &  53.14(16)  &  24 &  33.69(12)  &  36 &  13.52(10)  &  36 &  6.62(8) \\
14 &  53.41(16)  &  26 &  33.80(12)  &  39 &  13.55(10)  &  39 &  6.63(8) \\
15 &  53.65(16)  &  28 &  33.90(12)  &  42 &  13.57(10)  &  42 &  6.64(8) \\
16 &  53.87(16)  &     &             &     &   &   45 &  6.65(8) \\
17 &  54.06(17)  &     &             &     &   &   48 &  6.66(8) \\
18 &  54.24(17)  &     &             &     &   &   51 &  6.67(8) \\
19 &  54.41(17)  &     &             &     &   &   54 &  6.68(8) \\
20 &  54.56(17)  &     &             &     &   &   57 &  6.68(8) \\
21 &  54.70(17)  &     &             &     &   &      &          \\
\hline
\end{tabular}
\end{center}
\caption{Each column reports $10^6 \times a^4\chi_{\L}$ as a function of $n_\cool$ for a given $\beta$. Data refer to $N=5$.}
\label{tab:rawdata_N5}
\end{table}

\begin{table}[!htb]
\tiny
\begin{center}
\begin{tabular}{|cc|cc|cc|cc|}
\hline
\multicolumn{8}{|c|}{$N=6$}\\
\hline
\hline
\multicolumn{8}{|c|}{$10^6 \times a^4 \chi_{\L}(n_\cool)$}\\
\hline
\hline
\multicolumn{2}{|c}{$\beta=25.32$} &\multicolumn{2}{|c}{$\beta=25.70$}&\multicolumn{2}{|c}{$\beta=26.22$}&\multicolumn{2}{|c|}{$\beta=26.65$}\\
\hline
\hline
2  &  36.77(17)  &  2  &  19.83(9)   &  3  &  10.41(12)  &  3  &  5.51(7) \\
3  &  42.25(19)  &  4  &  23.82(11)  &  6  &  11.56(14)  &  6  &  6.07(8) \\
4  &  45.04(21)  &  6  &  25.25(12)  &  9  &  11.99(14)  &  9  &  6.28(8) \\
5  &  46.79(21)  &  8  &  26.03(12)  &  12 &  12.22(14)  &  12 &  6.39(8) \\
6  &  48.02(22)  &  10 &  26.52(12)  &  15 &  12.37(14)  &  15 &  6.47(8) \\
7  &  48.93(22)  &  12 &  26.87(12)  &  18 &  12.47(15)  &  18 &  6.52(8) \\
8  &  49.64(23)  &  14 &  27.13(13)  &  21 &  12.55(15)  &  21 &  6.55(8) \\
9  &  50.21(23)  &  16 &  27.33(13)  &  24 &  12.61(15)  &  24 &  6.58(8) \\
10 &  50.68(23)  &  18 &  27.50(13)  &  27 &  12.66(15)  &  27 &  6.61(8) \\
11 &  51.08(23)  &  20 &  27.64(13)  &  30 &  12.70(15)  &  30 &  6.63(8) \\
12 &  51.43(24)  &  22 &  27.75(13)  &  33 &  12.74(15)  &  33 &  6.64(8) \\
13 &  51.72(24)  &  24 &  27.85(13)  &  36 &  12.77(15)  &  36 &  6.66(8) \\
14 &  51.99(24)  &  26 &  27.94(13)  &  39 &  12.80(15)  &  39 &  6.67(8) \\
15 &  52.22(24)  &  28 &  28.02(13)  &  42 &  12.82(15)  &  42 &  6.68(8) \\
16 &  52.43(24)  &     &             &     &             &  45 &  6.69(8) \\
17 &  52.62(24)  &     &             &     &             &  48 &  6.70(8) \\
18 &  52.79(24)  &     &             &     &             &  51 &  6.70(8) \\
19 &  52.95(24)  &     &             &     &             &  54 &  6.71(8) \\
20 &  53.09(24)  &     &             &     &             &  57 &  6.72(8) \\
\hline
\end{tabular}
\end{center}
\caption{Each column reports $10^6 \times a^4\chi_{\L}$ as a function of $n_\cool$ for a given $\beta$. Data refer to $N=6$.}
\label{tab:rawdata_N6}
\end{table}

\FloatBarrier

\providecommand{\href}[2]{#2}\begingroup\raggedright\endgroup

\end{document}